\documentclass[letterpaper,11pt]{article}
\usepackage{amsmath,amssymb,yfonts,esint}

\usepackage{wrapfig}
\usepackage{floatflt}
\usepackage{mathrsfs}

\usepackage[pdftex]{color,graphicx}

\usepackage{dsfont}

\newcommand{\be}{\begin{equation}}
\newcommand{\ba}{\begin{eqnarray}}
\newcommand{\ea}{\end{eqnarray}}
\newcommand{\nn}{\nonumber}

\def\d{\delta}
\def\e{\epsilon}

\def\m{\mu}

\def\t{\tau}

\def\G{\Gamma}

\def\OO{\Omega}

\def\X{\Xi}

\def\ca{{\cal A}}
\def\cb{{\cal B}}

\def\cd{{\cal D}}

\def\ch{{\cal H}}

\def\co{{\cal O}}
\def\cp{{\cal P}}

\def\cu{{\cal U}}

\newcommand{\fe}{{\bf e}}

\newcommand{\fU}{{\bf U}}

\newcommand{\pa}{\partial}

\newcommand{\bbC}{{\Bbb C}}

\fontfamily{yfrak}

\title{From Quantum Gravity to Quantum Field Theory  \\  
 via Noncommutative Geometry}
\author{Johannes \textsc{Aastrup}$\,^{a}$\footnote{email: \texttt{aastrup@math.uni-hannover.de}}\;  \&
Jesper M\o ller \textsc{Grimstrup}\,$^{b}$\footnote{email: \texttt{grimstrup@nbi.dk}}\\[3ex]
$^{a}\,$\textit{Mathematisches Institut, Universit\"at Hannover,} \\ \textit{Welfengarten 1, 
D-30167 Hannover, Germany.}
\\[3ex]
$^{b}\,$\textit{The Niels Bohr Institute, University of Copenhagen,} \\   \textit{Blegdamsvej 17, DK-2100 Copenhagen, Denmark.}
}

\begin{document}
\maketitle
\begin{abstract}
A link between canonical quantum gravity and fermionic quantum field theory is established in this paper. From a spectral triple construction which encodes the kinematics of quantum gravity semi-classical states are constructed which, in a semi-classical limit, give a system of interacting fermions in an ambient gravitational field. The interaction involves flux tubes of the gravitational field. In the additional limit where all gravitational degrees of freedom are turned off, a free fermionic quantum field theory emerges.
\end{abstract}

\newpage
\tableofcontents

\section{Introduction}

In this paper we establish a link between canonical quantum gravity and fermionic quantum field theory. The starting point is a mathematical construction -- a {\it spectral triple} -- which utilizes Connes' noncommutative geometry to recombine central elements of canonical quantum gravity into a geometrical construction over a configuration space of general relativity \cite{AGN1, AGN2, AGN3}. 
From this spectral triple an infinite system of interacting fermions in an ambient gravitational field emerge in a semi-classical approximation. We discuss two different types of semi-classical states, one of which entail a fermionic interaction which involves flux-tubes of the gravitational field. 
Finally, a free fermionic quantum field theory emerges from the construction in the limit where all gravitational degrees of freedom are turned off.

The fermionic degrees of freedom encountered in the semi-classical approximation emerge as degrees of freedom related to a certain labeling of the semi-classical states. 
 The spectral triple construction itself involves, a priori, only gravitational degrees of freedom.\\

A spectral triple $(B,H,D)$ is the central ingredient in a generalization of Riemannian geometry known as noncommutative geometry. Its constituents are a $\star$-algebra $B$ represented as bounded operators on a Hilbert space $H$ together with an unbounded, self-adjoint operator $\cd$ called the Dirac operator. 
For a commutative algebra the spectral triple is, under certain requirements \cite{ConnesBook,Connes:1996gi}, equivalent to a Riemannian geometry on the state space of the algebra.  
The spectral triple which we investigate in this paper is a geometrical construction over a configuration space of connections. It is constructed over an infinite system of cubic lattices and the algebra is generated by loops running in these lattices. This algebra is inherently noncommutative. A copy of $SU(2)$ is assigned to each edge in these lattices - much alike lattice gauge theory - and with the refinement of lattices the number of copies of $SU(2)$ grows to infinity in a way which captures information of a configuration space of $SU(2)$ connections. This configuration space is related to canonical quantum gravity via Ashtekar variables \cite{Ashtekar:1986yd,Ashtekar:1987gu} and the algebra of loop represent holonomy transforms of the Ashtekar connection.
The Dirac type operator is then essentially given by a sum of Dirac operators on these copies of $SU(2)$. In \cite{AGN2, AGN3} it was proven that a semi-finite spectral triple emerge from these constituents, and in \cite{AGNP1} it was shown that the interaction between the algebra of loops and the Dirac type operator reproduces the structure of the Poisson bracket of general relativity formulated in terms of Ashtekar variables. This means that the spectral triple encodes information of the kinematics of quantum gravity.

The new insight presented in this paper is that a natural class of semi-classical states, labelled by a certain loop order, reside naturally in the Hilbert space associated to the spectral triple. In \cite{AGNP1,Aastrup:2010kb,Aastrup:2010ds} the states at first loop order were identified and were shown to entail the Dirac Hamiltonian in $3+1$ dimensions in a semi-classical approximation. Thus, these states were interpreted as one-particle states of a fermion in an ambient gravitational field dictated by the semi-classical approximation. Correspondingly, in this paper we find that the $n$'th order states entail, in a semi-classical approximation, a system of $n$ coupled fermions. In fact, we discuss two different types of semi-classical states. The first type, which we discuss in detail, give rise to a fermion interaction which involves flux tubes of the Ashtekar connection. The second type of states does not give rise to this interaction. 
 Finally, in the limit where all gravitational degrees of freedom are turned off, a free fermionic quantum field theory emerges. 
 The first type of states entail a certain {\it twisted} version of a free fermionic quantum field theory.  The 'twist' means that the inner product in what resembles a Fock space involves a certain mixing of spinor degrees of freedom.  If one restricts the construction to Weyl spinors this feature is absent and the results is an ordinary free fermionic quantum field theory. The second type of states which we discuss does not have this feature.

The construction of the semi-classical states follows a certain logic which takes its point of departure in the choice of algebra: the choice of a noncommutative algebra of loops necessitates the addition of a matrix factor in the Hilbert space to accommodate a representation of the algebra. Also, the algebra of loops comes with a dependency on a choice of basepoint needed in order to define a product between loops. To free the construction from this basepoint dependency one is lead to consider a type of states which {\it spread out the basepoint}. These states involve a certain matrix degree of freedom which will later merge into spinors. It is the additional matrix factor in the Hilbert space which permit these matrices to appear. Finally, the combination of these basepoint-independent states with coherent states over a classical point in the phase-space of gravity entails in a semi-classical approximation an infinite system of fermions which interact both among themselves and with the ambient gravitational fields given by the classical phase-space point. Thus, there is a direct line of reason from the choice of a noncommutative algebra of holonomy loops (as opposed, for example, to a commutative algebra of functions on the space of connections) to the emergence of fermionic degrees of freedom.\\

The paper is organized as follows: In section 2 we briefly introduce elements of noncommutative geometry and its relation to the standard model. In section 3 we introduce Ashtekar variables together with their dual, holonomy loops and fluxes of triad fields. In section 4 we then present the construction of the spectral triple. The triple is constructed first at the level of a single graph and second as a continuum limit over an infinite, countable set of embedded graphs. In section 5 we comment on the Poisson bracket of General Relativity which is encoded in the spectral triple through the interaction between the Dirac type operator and the algebra of loops. Then, with the basic construction presented, we are in section 6 ready to address the question regarding the semi-classical approximation. We start the analysis by reviewing first the issue of a certain basepoint dependency of the spectral triple. The resolution of this problem leads directly to states which, in a semi-classical limit, entail the Dirac Hamiltonian of a single particle. In section 7 we then show that also higher order states reside within the spectral triple which, in a semi-classical approximation, leads to a system of interacting fermions. We show that in the special limit where all gravitational degrees of freedom are turned off, a Fock space emerges. In section 8 
we give a conclusion. Detailed computations are given in appendix A.

\section{Noncommutative geometry}

Noncommutative geometry is based on the insight, due to Connes, that the metric on a compact Riemannian manifold can be recovered from the Dirac operator ${ D}$ and its interaction with the algebra of smooth functions on the manifold \cite{ConnesBook}. This means that the metric data is completely determined by the triple
\begin{equation}
(C^\infty(M),L^2(M,S),D)\;.
\label{NOET}
\end{equation}
This result entails a natural generalization of Riemannian geometry where one considers also noncommutative algebras. The central object in this generalization is the {\it spectral triple} $(A,H,D)$ where $A$ is a not necessarily commutative $*$-algebra, $H$ is a Hilbert space carrying a representation of $A$ and ${ D}$ is an unbounded, self-adjoint operator called the Dirac operator. Such a triple is called spectral if it satisfies two conditions:
\begin{enumerate}
\item
The resolvent of $\cd$, $(1+D^2)^{-1}$, is a compact operator in $H$.
\item
The commutator $[D,a]$ is bounded for all elements $a\in A$.
\end{enumerate}
A noncommutative geometry consist of a spectral triple which is required to satisfy an additional number of rules which generalize the interactions of the constituents in (\ref{NOET})  so that the construction coincides with Riemannian geometry whenever $A$ is commutative. These rules strongly restrict the choice of the Dirac operator ${ D}$.

It turns out that the standard model of particle physics coupled to the gravitational field provides an example of a noncommutative geometry. In this case the algebra is an {\it almost commutative algebra} of the form
\begin{equation}
A= C^\infty(M)\otimes A_F\;,\quad A_F= \mathbb{C}\oplus \mathbb{H}\oplus M_3(\mathbb{C})\;,
\label{AlG}
\end{equation}
which interacts with a Dirac operator that consist of two parts
$$
D= D_M + D_F\;,
$$
where $D_M$ is the standard Dirac operator on the manifold $M$ and $D_F$ is a matrix valued function on $M$ which encodes the metric data of the states over $A_F$. It was Connes who realized that the entire structure of the standard model coupled to general relativity is encoded in a spectral triple that involve this algebra and Dirac operator \cite{Connes:1996gi,Chamseddine:2006ep,Chamseddine:2007hz,Chamseddine:2007ia}. In fact, the abstract requirements for the Dirac operator, as mentioned above, entails that $D_F$ contains both the non-Abelian gauge fields and the Higgs field of the standard model together with their couplings to the elementary fermions. This very remarkable fact provides the Higgs field with a geometrical interpretation as a carrier of metric information on a noncommutative space. Furthermore, the action of the standard model coupled to the Einstein Hilbert action emerges from this spectral triple construction through the so-called spectral action principle \cite{Chamseddine:1991qh,Chamseddine:1996rw,Chamseddine:1996zu} which involves a heat-kernel expansion of the Dirac operator $D$.

What emerges from this spectral triple construction is essentially the classical action. Quantization of the fields of the standard model - barring the gravitational field - is applied {\it after} the heat kernel expansion. Thus, the work of Connes and Chamseddine raises the question whether quantum theory should not play a more prominent role in this approach to high-energy physics. Furthermore, since the construction of Connes and Chamseddine is fundamentally gravitational, one would expect that the answer to this question should, somehow, involve elements of quantum gravity. Thus, one might speculate that the spectral triple construction due to Connes should be interpreted as a low-energy limit of a theory of quantum gravity.

It was this line of reasoning that motivated the construction in  \cite{AGN1, AGN2, AGN3} of a semi-finite spectral triple over a configuration space of connections, see also \cite{Aastrup:2005yk}. There, the idea is to use noncommutative geometry to identify a natural non-perturbative construction within the framework of canonical quantum gravity and then, subsequently, identify a semi-classical limit which coincides with known physics. Ultimately, the goal is to make contact to Connes work on the standard model.

Before we present the construction of this semi-finite spectral triple we will briefly review Ashtekar variables and holonomy loops since these play a key role in the physical interpretation of the semi-finite spectral triple.

\section{Ashtekar variables and holonomy loops}
\label{canonicalgravity}


 Let $M$ be a 4-dimensional globally hyperbolic manifold 
and consider a foliation of $M$ according to $M = \mathbb{R}\times \Sigma$ where $\Sigma$ is a spatial manifold. Let $g_{mn}=e_m^a e_{na}$ be the corresponding spatial metric and $e_m^a$ the spatial dreibein. Here the letters $m,n, ...$ and $a,b,...$ denote curved and flat spatial indices.

The Ashtekar variables \cite{Ashtekar:1986yd,Ashtekar:1987gu} consist first of a complexified
$SU(2)$ connection $A_{m}^a(x)$ on $\Sigma$. The Ashtekar connection is a certain complex linear combination of the spatial spin connection and the extrinsic curvature of $\Sigma$ in $M$. The canonically conjugate variable to $A_{m}^a(x)$ is the inverse densitized dreibein
\[
{E}^m_a = e e_a^m\;,
\]
where $e=\mbox{det}( e_m^a)$. This set of variables satisfy the Poisson bracket
\[
\{ A_{m}^a(x),{E}_b^n(y)\} =  \kappa\d^a_b\d_m^n \d^{(3)}(x,y)\;,
\]
where $\kappa$ is the gravitational constant. The formulation of general relativity in terms of these variables involves an additional set of constraints, the Gauss, the Hamiltonian and the Diffeomorphism constraints. For detail we refer to \cite{AL1}.

The formulation of canonical gravity in terms of connection variables permits a shift to loop variables which are taken as the holonomy transform  
\begin{equation}
h_l(A)  = \cp \mbox{exp}\int_l A_m dx^m\;,
\label{hoool}
\end{equation}
along a loop $l$ in $\Sigma$. To define a conjugate variable to $h_l(A)$ let $dF_a$ be the flux of the triad field ${E}_a^m$ corresponding to an infinitesimal area element of the spatial manifold $\Sigma$, which can be written
\begin{equation}
dF_a = \e_{mnp} {E}^m_a dx^n\wedge dx^p \;.
\nonumber
\end{equation}
Given a 2 dimensional surface $S$ in $\Sigma$ the total flux of ${E}_a^m$ through $S$ reads
\begin{equation}
F_{S}^a= \int_S dF^a\;.
\label{fluuux}
\end{equation}
Next, consider a surface $S$ and let $l=l_1\cdot l_2$ be a line segment in $\Sigma$ which intersect $S$ at the point $l_1\cap l_2$. The Poisson bracket between the flux and holonomy variables read
\begin{equation}
\{ h_l, F_S^a \} =  \iota (S,l) \kappa h_{l_1}\t_a h_{l_2}\;.
\label{Poisson}
\end{equation}
where $\t$ denote the generators of the Lie algebra of $G$. Here, $\iota$ is given by
\[
\iota(S,l) = \pm 1, 0
\]
depending on the intersection between $S$ and $l$.

The holonomy and flux variables in (\ref{hoool}) and (\ref{fluuux}) are taken as the basic variables in loop quantum gravity and will also play a key role in the following. Here, however, we shall restrict ourselves to a real $SU(2)$ connection, known as the Ashtekar-Barbero connection. This restriction corresponds either to a Euclidian setting, in which the Hamilton constraint remains its simple form, or, alternatively, to a Lorentzian setting in which the Hamilton constraint requires an additional term. For detail we refer to \cite{AL1}.

\section{A spectral triple over holonomy loops}

In this section we outline the construction of the semi-finite spectral triple first presented in \cite{AGN1,AGN2} and further developed in \cite{AGN3}. The construction of the Dirac type operator follows \cite{AGN3}.

This spectral triple merges ideas and techniques of canonical gravity and noncommutative geometry. 
We first construct a spectral triple at the level of a finite graph. Next we take the limit of such spectral triples, over an infinite system of ordered graphs, to obtain a  limiting spectral triple.

The spectral triple construction works with a compact group $G$. This means that the construction does not a priori apply to the original Ashtekar connection, which takes values in complexified $SU(2)$, but rather to the Ashtekar- Barbero connection which takes values in $SU(2)$.

\subsection{Holonomy loops}

\begin{figure}[t]
\begin{center}
\resizebox{!}{3cm}{
 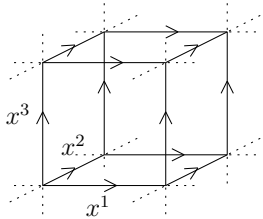}
\end{center}
\label{plaquet}
\caption{A plaquet in the lattice $\G$.}
\end{figure}
Let $\G$ be a 3-dimensional, finite, cubic lattice. Let $\{x_i\}$ and $\{l_j\}$ denote vertices and edges in $\G$, respectively. The edges in $\G$ are oriented according to the three main directions in $\G$, the $x^1$- $x^2$- and $x^3$-directions, see figure 1. Thus, an edge $l$ is a map
\[
l:\{0,1\}\rightarrow \{x_i\}\;,
\]
where $l(0)$ and $l(1)$, the start and endpoints of $l$, are adjacent vertices in $\G$. A sequences of edges $\{ l_{i_1}, l_{i_2},\ldots, l_{i_n}\}$ where $l_{i_j}(1)= l_{i_{j+1}}(0)$ is a based loop if $l_{i_1}(0)=l_{i_n}(1)=x_0$ where $x_0\in \{x_i\}$ is a preferred vertex in $\G$ called the basepoint. An edge has a natural involution given by reversing its orientation. Thus, 
\[
l_i^*(t)=l_i(1-t)\;,
\]
and the involution of a loop $L=\{ l_{i_1}, l_{i_2},\ldots, l_{i_n}\}$ is given by
\[
L^*=\{ l^*_{i_n}, \ldots,l^*_{i_2}, l^*_{i_1}\}\;.
\]
In the following we shall discard trivial backtracking which means that we introduce the equivalence relation
\[
\{\ldots,l_{i_{j-1}}, l_{i_j}, l^*_{i_j},\ldots\} \sim \{\ldots, l_{i_{j-1}},\ldots\}\;,
\]
and let a loop $L$ be an equivalence class with respect hereto. 

The product between two loops $L_1=\{l_{i_j}\}$ and $L_2=\{l_{i_l}\}$ is simply given by gluing the loops to form a new sequence of edges:
\[
L_1\cdot L_2 = \{l_{i_1},\ldots,l_{i_n},l_{k_1},\ldots,l_{k_m}\}\;.
\]
One easily checks that the involution equals the inverse which gives the set of loops in $\G$ the structure of a group, known as the hoop group.

Finally, we consider finite series of loops
\begin{equation}
a =\sum_i a_i L_i\;,\quad a_i\in \mathbb{C}\;,
\label{form}
\end{equation}
with the involution
\[
a^* = \sum_i  \bar{a}_i L^*_i\;,
\]
and the product  between $a$ and a second element $b=\sum_j b_j L_j$ 
\[
a \cdot b = \sum_{i,j} (a_i b_j) L_i\cdot L_j \;.
\]
The set of elements of the form (\ref{form}) is a $\star$-algebra denoted $\cb_\G$.

\subsection{ Generalized connections}

Next, let $G$ be a compact, connected Lie-group.  For the aim of this paper it is natural to choose $G=SU(2)$. We shall, however, develop the formalism for  general groups. Let $\nabla$ be a map
\[
\nabla:\{l_i\}\rightarrow G\;,
\]
which satisfies
\[
 \nabla(l_i) = \nabla(l_i^*)^{-1}\;,
\]
and denote by $\ca_\G$ the set of all such maps. Clearly,
\[
\ca_\G \simeq G^{n(\G)}\;,
\]
where the total number of vertices in $\G$ is written $n({\G})$. 
Given a loop $L=\{ l_{i_1}, l_{i_2},\ldots, l_{i_n}\}$ let
\[
\nabla(L)= \nabla(l_{i_1})\cdot \nabla(l_{i_2})\cdot \ldots\cdot \nabla(l_{i_n})\;.
\]
This turns $\nabla$ into a homomorphism from the hoop group into $G$ and provides a norm on $\cb_\G$
\[
\parallel a\parallel = \sup_{\nabla\in\ca_\G}\parallel\sum_i a_i\nabla(L_i)\parallel_{G}\;,\quad    a\in\cb_{\G}\;,
\]
where the norm on the rhs is the matrix norm given by a chosen representation of $G$. The closure of the $\star$-algebra of loops with respect to this norm is a $C^\star$-algebra\footnote{Note that the natural map from $\cb_\G$ to $B_\G$ is not necessarily injective.}. We denote this loop algebra by $B_\G$.

\subsection{A spectral triple over $\ca_\G$}

First, let $\ch_\G$ be the Hilbert space
\begin{equation}
L^2(G^{n(\G)},Cl(T^*G^{n(\G)})\otimes M_l(\mathbb{C}))\;,
\label{jay}
\end{equation}
where $L^2$ is with respect to the Haar measure and where $l$ is the size of the matrix representation of $G$. Here, $Cl(T^*G^{n(\G)})$ is the Clifford bundle of the cotangent bundle over $G^{n(\G)}$ with respect to a chosen left and right invariant metric.  There is a natural representation of the loop algebra on $\ch_\G$ given by
\[
f_L \cdot \Psi(\nabla)= (1\otimes \nabla(L))\Psi(\nabla)\;,\quad \Psi\in \ch_\G\;,
\]
where the first factor acts on the Clifford bundle and the second factor acts on the matrix factor in $\ch_\G$.

Next, denote by $\cd_\G$ a Dirac operator on $\ca_\G$. The precise form of $\cd_\G$ will be determined below through the process of taking the limit over graphs. 
In total, the triple $(B_\G,\ch_\G,\cd_\G)$  is a spectral triple associated to the graph $\G$.

\subsection{The limiting spectral triple}

\begin{figure}[t]
\begin{center}
\resizebox{!}{2.5cm}{
 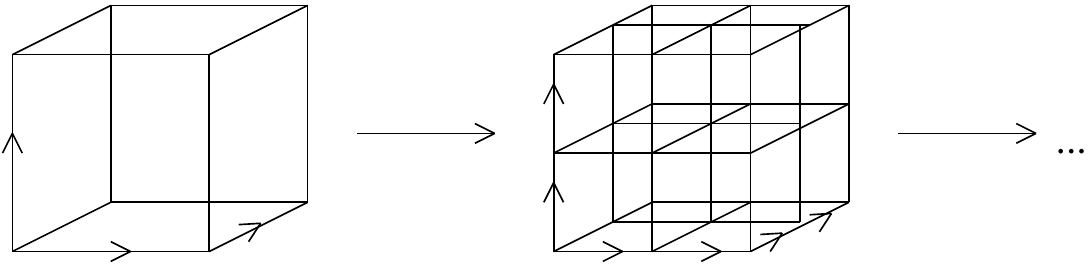}
\end{center}
\caption{Subdivision of a cubic lattice cell into 8 new cells.}
\end{figure}
The goal is to obtain a spectral triple over the space $\ca$. To do this we take the limit of spectral triples over the intermediate spaces $\ca_{\G}$.

Let $\{\G_i\}$, $i\in \mathbb{N}$, be an infinite sequence of 3-dimensional, finite, cubic lattices where $\G_{i+1}$ is the lattice obtained from $\G_i$ by subdividing each elementary cell in $\G_i$ into 8 new cells. This process involves the subdivision of each edge $l_j$ in $\G_i$ into two new edges in $\G_{i+1}$ together with the addition of new vertices and edges, see figure 2. We denote the initial lattice by $\G_0$.
Corresponding to this sequence of cubic lattices there is a projective system $\{\ca_{\G_i}\}$ of spaces obtained from the graphs $\{\G_i\}$, together with natural projections between these spaces
\begin{equation}
P_{i,i+1}: \ca_{\G_{i+1}}\rightarrow \ca_{\G_{i}}\;.
\label{structure}
\end{equation}
Consider now a system of triples
\[
(B_{\G_i},\ch_{\G_i},\cd_{\G_i})\;,
\]
with the restriction that these triples are compatible with the projections (\ref{structure}). This requirement is easily satisfied for the algebras and the Hilbert spaces, see \cite{AGN3}. For the Dirac type operators the problem reduces to the simple case where an edge in $\G_i$ is subdivided into two edges in $\G_{i+1}$, see figure 3.1. This corresponds to the projection
\begin{figure}[t]
\begin{center}
\resizebox{!}{3cm}{
 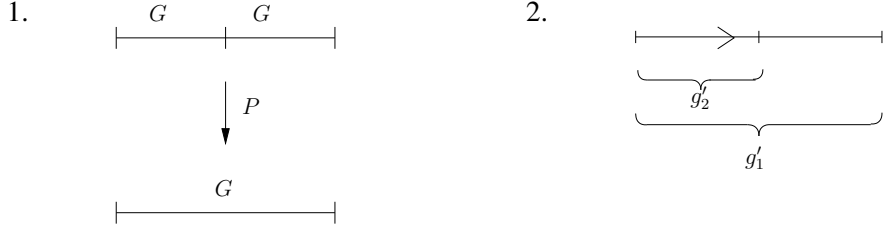}
\end{center}
\caption{A subdivision of an edge into two and the new parameterization of the edge.}
\end{figure}
\begin{equation}
P: G^2 \rightarrow G\;,\quad (g_1,g_2)\rightarrow g_1\cdot g_2\;,
\label{firstproj}
\end{equation}
and a corresponding map between Hilbert spaces
$$
P^*: L^2(G,Cl(T^*G)\otimes M_l)\rightarrow L^*(G^2,Cl(T^*G^2)\otimes M_l)\;.
$$
The compatibility condition for the Dirac type operator reads
\begin{equation}
P^* (\cd_{1} v)(g_1,g_2) = \cd_{2} (P^*v)(g_1,g_2) \;,\quad v\in L^2(G,Cl(T^*G)\otimes M_l)\;.
\nonumber
\end{equation}
Here $\cd_{1}$ is the Dirac operator on $G$,  and $\cd_{2} $ is the corresponding Dirac operator on $G^{2}$. 

Consider the following change of variables
\begin{equation}
\Theta: G^2\rightarrow G^2\; ;   (g_1,g_2)\rightarrow (g_1\cdot g_2,g_1)=: (g_1',g_2')\;,
\label{change}
\end{equation}
for which projection (\ref{firstproj}) obtains the simple form
\begin{equation}
P(g_1',g_2') = g_1'\;.
\label{proJ2}
\end{equation}
This change of variables corresponds to a new parameterization of the edge, see figure 3.2.
It is now straight forward to write down a Dirac operator on $G^2$ which is compatible with the projection (\ref{proJ2}). Basically, we can pick any Dirac operator of the form
\begin{equation}
\cd_2  = \cd_1 + a \cd'_2\;,\quad a\in \mathbb{R}\;,
\nonumber
\end{equation}
where $\cd'_2$ is a Dirac operator on the copy of $G$ in $G^2$ whose coordinates are eliminated by the projection (\ref{proJ2}). At this point the choice of the operator $\cd'_2$ is essentially unrestricted with $a$ being an arbitrary real parameter. However, for reasons explained in \cite{AGN3} it turns out that  $\cd_1$ and $\cd'_2$ should of the form
\begin{equation}
\cd_i = \sum_j    {\bf e}_i^j \cdot L_{{\bf e}_i^j}\;,
\label{dir}
\end{equation}
where the product is Clifford multiplication. In equation (\ref{dir}) $\{e_i^j\}$ denotes a left-translated orthonormal basis of $T^*G$ where $G$ is the $i$'th copy in $G^n$. $L_{e_i^j}$ denotes the corresponding differential.

This line of analysis is straightforwardly generalized to repeated subdivisions. At the level of the $n$'th subdivision of the edge the change of variables which generalizes (\ref{change}) reads
\begin{eqnarray}
\Theta:G^n  \rightarrow G^n\; ; &&
\nn\\ (g_1,g_2,\ldots,g_n)&\rightarrow& (g_1\cdot g_2 \cdot \ldots\cdot g_n, g_2 \cdot \ldots\cdot g_n, \ldots, g_n)
\nn\\&:=&(g_1',g_2',\ldots,g_n')
\label{genchange}
\end{eqnarray}
which corresponds to the structure maps
\[
P_{n,n/2}: G^n \rightarrow G^{n/2}\; ; \quad (g_1',g_2',g_3',\ldots,g_n') \rightarrow (g_1',g_3',\ldots,g_{n-1}')\;.
\]
Again, it is straightforward to construct a Dirac type operator compatible with these structure maps. This construction gives rise to a series of free parameters $\{a_i\}$, one for each subdivision. 
Thus, by solving the $G^2\rightarrow G$ problem repeatedly, and by piecing together the different edges, we end up with a Dirac type operator on the level of $\Gamma_n$ of the form
\begin{equation}
\cd_{\G_n}=\sum_i a_i \cd_i\;,
\label{general}
\end{equation}
where $\cd_i$ is a Dirac type operator corresponding to the $i$'th level of subdivision in $\ca_{\G_n}$.

The change of variables in (\ref{genchange}) is the key step to construct $\cd_{\G_n}$. However, there will be many different partitions of the line segment which simplify the structure maps and lead to different Dirac type operators. This ambiguity was commented on in \cite{AGNP1}. We refer to \cite{AGNP1} for a thoroughly discussion which concludes that a single type of subdivision stand out as "natural" due to the classical interpretation of the corresponding Dirac type operator.\\

We are now ready to take the limit of the triples $(B_{\G_i},\ch_{\G_i},\cd_{\G_i})$. First, the Hilbert space $\ch$ is the inductive limit of the intermediate Hilbert spaces $\ch_{\G_i}$. That is, it is constructed by adding all the intermediate Hilbert spaces 
$$
\ch' =\oplus_{\G\in \{\G_i\}}L^2 (G^{n(\G)},Cl(T^*G^{n(\G)})\otimes M_l({C}))/N\;, 
$$
where $N$ is the subspace generated by elements of the form 
$$
(\ldots , v, \ldots , -P^*_{ij}(v),\ldots )\;,
$$
where $P^*_{ij}$ are the induced maps between Hilbert spaces. The Hilbert space $\ch$ is then the completion of $\ch'$. The inner product on $\ch$ is the inductive limit inner product. This Hilbert space is manifestly separable. 
Next, the algebra
$$
\cb:= \lim_{\stackrel{\G}{\longrightarrow}}\cb_{\G}\;
$$
contains loops defined on a cubic lattice $\G_n$ in $\{\G_n\}$. 
Note that the algebra $\cb$ differs from the algebra used in loop quantum gravity on two points: first, we only consider loops running in cubic lattices, whereas the algebra in loop quantum gravity is generated by piece-wise analytic loops. Second, we consider loops which corresponds to untraced holonomy loops. Thus, the algebra $\cb$ is noncommutative.

Finally, the Dirac-like operator $D_{\G_n}$ descends to a densely defined operator on the limit Hilbert space $\ch$
$$
\cd = \lim_{\stackrel{\G}{\longrightarrow}}\cd_{\G_n}\;.
$$
We factorize $\ch'$ in 
$$\lim L^2(G^{n(\Gamma)}, M_l)\otimes \lim Cl(T^*_{id}(G^{n(\Gamma)})).$$ 
On $\lim Cl(T^*_{id}(G^{n(\Gamma)}))$ there is an action of the algebra $\lim Cl(T^*_{id}(G^{n(\Gamma)}))$. The completion of this algebra with respect to this action is the CAR algebra and admits a normalized trace, i.e. $tr(1)=1$. Let $Tr$ be the ordinary operator trace on the operators on $\lim L^2(G^{n(\Gamma)}, M_l)$ and define $\tau =Tr\times tr$. In \cite{AGN2,AGN3} we prove that for $G=SU(2)$ the triple $(\cb,\ch,\cd)$ is a semi-finite spectral triple with respect to $\tau$  when the sequence $\{a_n\}$ converges to infinity. This means that:
\begin{enumerate}
\item
$
(1+\cd^2)^{-1}
$
is $\tau$-compact, i.e. can be approximated in norm with finite trace operators, and
\item
the commutator $[\cd,a]$ is bounded.
\end{enumerate}
In \cite{Lai:2010ig} this result has been proven to hold for a general compact Lie group.

\section{Link to canonical quantum gravity}
\label{ThePoissonStructure}

This spectral triple encodes information about the kinematics of quantum gravity.
Consider first the spaces $\ca_{\G_i}$ and their projective limit.
Denote by
\[
\overline{\ca} := \lim_{\stackrel{\G}{\longleftarrow}}\ca_\G\;.
\]
Further, given a trivial principal $G$-bundle denote by $\ca$ the space of all smooth connections herein. In \cite{AGN3} we proved that $\ca$ is densely embedded in $\overline{\ca} $:
\[
\ca \hookrightarrow \overline{\ca} \;.
\]
This fact justifies the terminology {\it generalized connections} for the completion $\overline{\ca}$ and shows that the semi-finite spectral triple $(\cb,\ch,\cd)$ is indeed a geometrical construction over the space $\ca$ of smooth connections.

Next, in \cite{AGNP1} we have shown that the interaction between the Dirac type operator $\cd$ and the algebra $\cb$ in the semi-finite spectral triple $(\ch,\ch,\cd)$ reproduces the structure of the Poisson bracket (\ref{Poisson}). 

What we find is the following: consider the $n$'th level of subdivision of an edge $l_i$ and consider a vector field $L_{{\bf e}^a_{j}} $ which corresponds to a copy of $G$ assigned to a segment of $l_i$ that emerges at this level of subdivision. Then $L_{{\bf e}^a_{j}} $ corresponds to a quantization of a flux variable sitting at the endpoint of this segment 
\begin{equation}
F^a_{\Delta S_{j}} \stackrel{\mbox{\tiny quantization}}{\longrightarrow} l_P^2 L_{{\bf e}^a_{j}} +  l_P^2 \OO^a_{k} \;,
\label{xxx}
\end{equation}
where $\OO^a_{k}$ is a correction term that consist of a certain combination of twisted, right-invariant vector fields acting on the copies of $G$ assigned to segments of $l_i$ which are situation "higher" in the inductive system of lattices. Put differently,  $\OO^a_{k}$ probes information which is more coarse grained relative to the line segment to which the $(i+s)$'th copy of $G$ is assigned. In (\ref{xxx}) $\Delta S$ refers to a surface located at the endpoint of the specific line segment, with the size\footnote{This size refers to the coordinate system which emerges from the projective system of cubic lattices. Thus, an edge which appears in the graph $\G_0$ is of length '1'.} $2^{-2n}$. For details we refer to \cite{AGNP1}.

In the following we shall ignore the correction terms $\OO^a_{k}$ when we apply relation (\ref{xxx}) to translate quantized quantities involving the Dirac type operator $\cd$ to their classical counterparts. The reason for this will become clear in the next section where we construct semi-classical states. These states have the property that any dependency on finite parts of the inductive system of lattices vanishes in a combined semi-classical limit continuum limit.

In the limit of repeated subdivision of lattices we find that the semi-finite spectral triple $(\cb,\ch,\cd)$ encodes information tantamount to a representation of the Poisson bracket of general relativity. Thus, the triple captures information about the kinematical sector of quantum gravity.

\section{A semi-classical approximation and a continuum limit}

In this section we review and further develop the analysis of semi-classical states found in $\ch$. The relevant references are \cite{AGNP1,Aastrup:2010ds}. In the following we set $G=SU(2)$ and choose a two-by-two matrix representation hereof (thus, in (\ref{jay}) we set $l=2$). 

The semi-classical analysis comes together with a certain continuum limit in which we discard information related to finite graphs. Thus, in this limit we 'zoom in' on infinitesimal edges only.

\subsection{Dependency on the choice of basepoint}
\label{dependency}

The semi-classical analysis starts with the realization that the spectral triple construction comes with a dependency on the choice of basepoint $x_{0}$. This was first pointed out in \cite{Aastrup:2010kb}. Had we instead chosen to work with traced holonomy loops this dependency would not show up since the basepoint dependency vanishes due to the cyclicity of the trace. 
In \cite{Aastrup:2010ds} this observation was taken as the point of departure for a construction which ultimately entail the emergence of the Dirac Hamiltonian for a single fermion and - as we shall see in the following - for a system of interacting fermions, in a semi-classical approximation. 

Consider first the graph $\G_n$ and an edge $l_i$ in $\G_n$. Associated to $l_i$ the operator
$$
{\bf U}_i:=  \frac{\mathrm{i} }{2}\left(   {\bf e}_i^a  \sigma^a  + {\bf e}_i^1{\bf e}_i^2{\bf e}_i^3 \right) g_i
$$
and check that ${\bf U}_i^*{\bf U}_i={\bf U}_i {\bf U}_i^*=\mathds{1}_2$ (see appendix A). Here, $g_i=\nabla(l_i)$ is an element in the copy of $G$ assigned to the edge $l_i$. 
Given an element $a\in\cb_{\G_n}$ we compute
\begin{equation}
\mbox{Tr}_{\tiny Cl}\left(  {\bf U}_i a {\bf U}_i^* \right) =  a_0 \mathds{1}_2  \;,
\label{Trace}
\end{equation}
where we write $a= a_0 \mathds{1}_2  + a^i \sigma^i$ with $\sigma^i$ being the Pauli matrices, and where $\mbox{Tr}_{\tiny Cl}$ denotes the trace over the Clifford algebra. Thus, conjugating with ${\bf U}_i$ singles out the matrix trace of $a$. Next, let $p=\{ l_{i_1}, l_{i_2},\ldots , l_{i_k}  \}$ be a path in $\G_n$ and define the associated operators by
\begin{equation}
{\bf U}_p := {\bf U}_{i_1}{\bf U}_{i_2} \ldots {\bf U}_{i_k}  \;,\qquad {U}_p := \nabla(l_{i_1})\cdot \nabla(l_{i_2})\cdot \ldots \nabla(l_{i_k})\;,
\label{twooperators}
\end{equation}
where $U_p$ is the ordinary parallel transport along $p$. The operators ${\bf U}_p$ form a family of mutually orthogonal operators labelled by paths in $\G_n$
\begin{equation}
 \mbox{Tr}_{\tiny Cl}\left(  {\bf U}_p^* {\bf U}_{p'}\right) = \d_{p,p'}\;.
\nn
\end{equation}
This relation\footnote{This orthogonality relation is only conditionally correct. If $p=p^{-1}$ or if $p$ is a loop which runs through its course twice, then additional factors arises, see formula (\ref{skidtpar}) for details on the relevant computations.} is due to the presence of the Clifford algebra elements in ${\bf U}_p$. Here $\d_{p,p'}$ equals one when the paths $p$ and $p'$ are identical and zero otherwise.

Consider now states in $\ch_{\G_n}$ of the form
\begin{equation}
\Psi_{n}(\psi) = 2^{-3n}\sum_i {\bf U}_{p_i}\psi(x_i) U^{-1}_{p_i}\;,
\label{sss}
\end{equation}
where the sum runs over vertices $x_i$ in $\G_{n}$ and where the path $p_i$ connects the basepoint $x_0$ with vertices $x_i$. Here $\psi(x_i)$ denotes an element in $M_2(\mathbb{C})$ associated to the vertex $x_i$. Later $\psi(x_i)$ will be seen to form a spinor degree of freedom at the point $x_i$.

For the purpose of this section the sum in (\ref{sss}) may run over all vertices. Later, however, it will be necessary to restrict this sum to a subclass of vertices. First, consider {\it edges} in $\G_n/\G_{n-1}$, that is, edges which lie in $\G_n$ but not in $\G_{n-1}$. 
To each edge is associated two vertices, and therefore two different paths, where one is a single "step" longer than the other. Let us denote them $ {p_i}$ and $ {p_{i+1}}$ . When we consider the operators $ {\bf U}_{p_i}$ and $ {\bf U}_{p_{i+1}}$  connecting the basepoint with such pairs of vertices, then we shall only consider pairs where the {\it shortest} path $p_i$ corresponds to an {\it even} operator $ {\bf U}_{p_i}$ ('even' is with respect to the Clifford algebra). Thus, we let the sum in (\ref{sss}) run over these vertices. This will be important when we develop the semi-classical analysis.

Now, $\Psi_n$ is a state in $\ch_{\G_n}$ which does not show the dependency on the choice of basepoint $x_0$ mentioned above. This means that the expectation value of an element $a$ in $\cb_{\G_n}$ on this state will depend only on the trace of $a$
\begin{equation}
\langle \Psi_{n} \vert a \vert \Psi_{n}\rangle =\langle \Psi_{n} \vert  \mbox{Tr}(a) \vert \Psi_{n}\rangle = \langle   \mbox{Tr}(a)\rangle\sum_i \psi^*(x_i)\psi(x_i)\;,
\label{indy}
\end{equation}
and since the trace of an element of $\cb_{\G_n}$ is basepoint independent, these states circumvent this problem.

\subsection{Coherent states in $\ch$}

The strategy is to combine states of the form (\ref{sss}) with a semi-classical approximation. To do this we
need to introduce coherent states in $\ch$. We first recall results for coherent states on various copies of $SU(2)$. This construction uses results of Hall \cite{H1,H2} and is inspired by the articles \cite{BT1,TW,BT2}.

First pick a point $(A_n^a,E^m_b)$ in the phase space of Ashtekar variables\footnote{recall that we work here with a real $SU(2)$ connection.} on a 3-manifold $\Sigma$. The states which we construct will be coherent states peaked over this point.
Consider first a single edge $l_i$ and thus one copy of $SU(2)$. Let $\{{\bf e}^a_i\}$ be a basis for $\mathfrak{su}(2)$. 
There exist families $\phi^t_{l_i}\in L^2(SU(2))$ such that
\begin{equation}
 \lim_{t \to 0}\langle \phi^t_{l_i}, t L_{{\bf e}^a_i}\phi_{l_i}^t \rangle=2^{-2n}\mathrm{i}E_a^m(x_{j})\;,
 \label{JP1}
 \end{equation}
and
\begin{equation}
\lim_{t \to 0}\langle \phi_{l_i}^t\otimes v, \nabla(l_i)\phi_{l_i}^t\otimes v \rangle=(v,h_{l_i}(A)v)\;,
\label{JP2}
\end{equation}
where $v \in \bbC^2$, and $(,)$ denotes the inner product hereon; $x_{j}$ denotes the 'right' endpoint of $l_i$ (we assume that $l_i$ is oriented to the 'right'), and the index "$m$" in the $E^m_a$ refers to the direction of $l_i$. The factor $2^{-2n}$ is due to the fact that $L_{{\bf e}_j^a}$ corresponds to a flux operator with a surface determined by the lattice \cite{AGNP1}.
Corresponding statements hold for operators of the type $$f(\nabla(l_i))P(t L_{{\bf e}^1_i},t L_{{\bf e}^2_i},t L_{{\bf e}^3_i}),$$  where $P$ is a polynomial in three variables, and $f$ is a smooth function on $SU(2)$, i.e.
$$ \lim_{t \to 0}\langle \phi^t_{l_i} f(\nabla(l_i))P(t L_{{\bf e}^1_i},t L_{{\bf e}^2_i},t L_{{\bf e}^3_i}) \phi^t_{l_i} \rangle=f(h_{l_i}(A))P(\mathrm{i}E_1^m,\mathrm{i}E_2^m,\mathrm{i}E_3m)\;.$$

 The states $\phi^t_{l_i}$  have further important physical properties which we are however not going to use at the present stage of the analysis. Also, the precise construction of these states, in particular the choice of complexifier \cite{TW}, is irrelevant for the results presented in this paper.

Let us now consider the graph $\G_n$. We split the edges into $\{ l_i \}$, and $\{ l'_i\}$, where $\{ l_i\}$ denotes the edges appearing in the $n$'th subdivision but not in the $n-1$'th subdivision, and $\{ l'_i \}$ the rest. Let
$\phi^t_{l_i}$ be the coherent state on $SU(2)$ defined above and
define the states $\phi_{l'_i}$ by
$$\lim_{t \to 0}\langle \phi^t_{l'_i} \otimes v ,\nabla(l_i) \phi^t_{l'_i}\otimes  v \rangle =  (v,h_{l'_i}(A)v)\;,$$
and
$$\lim_{t \to 0}\langle \phi^t_{l'_i} , t L_{{\bf e}_j^a} \phi^t_{l'_i}\rangle = 0\;.$$
Finally define $\phi^t_n$ to be the product of all these states as a state in $L^2(\ca_{\Gamma_n})$. These states are essentially identical to the states constructed in \cite{TW} except that they are based on cubic lattices and a particular mode of subdivision.

In the limit $n\rightarrow\infty$ these states produce the right expectation value on all loop operators in the infinite lattice. The reason for the split of edges in $l_i$ and $l'_i$ in the definition of the coherent state is to pick up only those degrees of freedom which 'live' in the continuum limit $n\rightarrow\infty$. In this way we shall, once the continuum limit is taken, partially have eliminated dependencies on finite parts of the lattices. In a classical setup, this amounts to information which has measure zero in a Riemann integral.

\subsection{The Dirac operator in 3 dimensions}

We are now ready to combine the content of the previous two subsections.
 Consider first the states
\begin{equation}
\Psi^t_n(\psi) :=  \Psi_{n}(\psi)  \phi^t_n
\label{dadada}
\end{equation}
composed by the states (\ref{sss}) and the coherent states introduced in the previous section. Furthermore, we will restrict the series $\{a_j\}_{j\in\mathbb{N}_+}$ of parameters in $\cd$: we now require all $a_j$'s associated to edges appearing in the $k$'th subdivision but not in the $k-1$'th subdivision to be equal. With this restriction the parameters $a_j$ represent a scaling degree of freedom.

We find 
\begin{eqnarray}
\langle \Psi^t_n\vert  \cd \vert \Psi^t_n\rangle 
 &=&
 \frac{a_n}{2^{5n+1}} 
 \sum_i \Big(  \psi^*(x_i) \sigma^a  E^m_a g_{i} \psi(x_{i+1}) g_i^{-1} 
 +( \sigma^a g_{i}    {\psi}(x_{i+1})g_i^{-1} )^*  E^m_a \psi(x_{i}) 
 \Big)\;,
 \label{hmmmmm}
\end{eqnarray}
where $x_i$ and $x_{i+1}$ denote start and endpoint of an edge $l_i$ and where the sum runs over edges appearing in the $n$'th but not in the $n-1$'th subdivision. 
The evaluation over coherent states has been performed, see equation (\ref{JP1}) and (\ref{JP2}), and therefore $E^m_a$ appears in (\ref{hmmmmm}) where the index '$m$' refers to the direction of the edge $l_i$.
Expand $g_{i}$ according to
$$
g_{i}= \mathds{1}_2 + \e A_m(x_i) +\co(\e^2)\;,
$$
where "$m$" is again the direction of the edge  $l_i$. This expansion is permitted whenever we apply the coherent states and take the continuum limit $n\rightarrow\infty$ together with the semi-classical limit $t\rightarrow 0$. 
The term $ g_{i} \psi(x_{i+1}) g^{-1}_{i} $, in the combined semi-classical and continuum limit, give a covariant derivative
 \begin{eqnarray}
\lim_{n\rightarrow\infty}\lim_{t\rightarrow 0}    \left(  g_{i+1} \psi (x_{i+1}) g^{-1}_{i} \right) &=&   ( \mathds{1}_2  + \e A_m )(\psi(x) + \e\partial_m\psi(x)) ( \mathds{1}_2  - \e A_m ) +\co(\e^2)  
\nn\\
&=& \psi(x) + \e \nabla_m\psi(x) +\co(\e^2) 
\label{zeroth}
\end{eqnarray}
where '$m$' denotes the direction of the edge $l_i$  and where $\nabla_m=\pa_m+ [A_m,\cdot]$ is the covariant derivative.
Finally, from equation (\ref{hmmmmm}) we obtain the limit
\begin{equation}
\lim_{n\rightarrow\infty}\lim_{t\rightarrow 0} \langle \Psi^t_n\vert t \cd \vert \Psi^t_n\rangle =
 \frac{1}{2}  \int_\Sigma d^3x  {\psi}^*(x) \Big( \sigma^a  E^m_a \nabla_m +  \nabla_m \sigma^a E^m_a     \Big) \psi(x) \;,
 \label{hmmm}
\end{equation}
provided we set $\e=2^{-n}$ and fix the parameters $\{a_n\}$ to $a_n= 2^{3n}$.

Thus, the expectation value of the Dirac type operator $\cd$ on the states (\ref{dadada}) renders, in a combined semi-classical and continuum limit, the expectation value of an ordinary spatial Dirac operator on a manifold $\Sigma$, on the condition that we fix the free parameters $a_n$ which appear in $\cd$. The spinor degrees of freedom emerged from the matrix factor which was introduced in the Hilbert space $\ch$ in order to accommodate a representation of the algebra of loops.

 In the computation leading to (\ref{hmmm}) it is crucial that the terms from (\ref{zeroth}) at zero'th order in $\e$ cancel out. If they had failed to do so the result would be terms which diverge in the continuum limit. This issue becomes particularly urgent when we in the next section consider many-particle states. There the grading of certain paths will give rise to signs which jeopardize this cancellation of zero order terms and forces us to impose certain conventions, see the paragraph after (\ref{supermanN}).

The expression (\ref{hmmm}) is formulated with respect to a certain coordinate system which emerges from the graphs $\G_n$. Thus, the cubic lattices can, in this specific limit, be interpreted as an emergent coordinate system. In particular, this coordinate system is exactly the coordinate system in which the classical Ashtekar variables are formulated\footnote{Of course, the Ashtekar variables do not depend on a particular choice of a coordinate system. However, the way they enter the analysis in this paper, they are written down with respect to a coordinate system.}. 

Notice that in order to obtain expression (\ref{hmmm}) we had to fix the parameter $a_n$. Thus, the freedom which we encountered when we constructed the Dirac type operator $\cd$ is eliminated in order to obtain a sensible semi-classical limit.

Notice also that $E^m_a$ is the densitized dreibein $E^m_a = \sqrt{g} e^m_a $ where $e^m_a$ is the dreibein and $g$ is the determinant of the 3-metric $g$ given by the dreibein. This means that the inverse measure $d^3x \sqrt{g}$ appear naturally in (\ref{hmmm}). However, a problem regarding the normalization of spinors does arise in this framework, see \cite{AGNP1}.

\subsection{The Dirac Hamiltonian}

In order to obtain the Dirac Hamiltonian instead of the Dirac operator in (\ref{hmmm}) we must introduce additional degrees of freedom which, in the appropriate limit, can correspond to the lapse and shift fields. This will then encode what amounts to a choice of foliation of space-time. These degrees of freedom are, seen from the spectral triple construction, related to the relationship between the Clifford algebra and the matrix factor in $\ch$. There are alternative ways to introduce these additional degrees of freedom, see \cite{Aastrup:2010kb,Aastrup:2009dy}, and at present it is not clear which approach is more natural.  For instance, following \cite{Aastrup:2009dy}

we can introduce a modification of the Dirac operator (\ref{general}) with a matrix factor associated to each edge, written
\begin{equation}
\label{modifications}
\cd_M = \sum_{i,a} a_i M_i {e}_i^a L_{e_i^a}
\end{equation}
where $M_i$ is an arbitrary, self-adjoint two-by-two matrix associated to the edge $l_i$. In the sum in (\ref{modifications}) $k$ runs over the different copies of the group according to the change of variables introduced in section (4.4) and $a$ is an $SU(2)$ index. One finds that the expectation value of $\cd_M$ on the states (\ref{dadada}) gives
\begin{eqnarray}
\lim_{n\rightarrow\infty}\lim_{t\rightarrow 0} \langle {\Psi}^t_n\vert t \cd_M \vert {\Psi}^t_n\rangle
&=& 
 \frac{1}{2}  \int_\Sigma d^3x  {\psi}^*(x) \Big( (N + N^b \sigma^b)(\sigma^a  E^m_a \nabla_\m  + \nabla_\m  \sigma^a  E^m_a  )   \Big) \psi(x)
 \nn\\
 && +\;\; \mbox{zero order terms}
  \;,
 \label{hmmmdi}
\end{eqnarray}
where we wrote $M_i$ as $N(x)\mathds{1}_2  + \mathrm{i}N^a(x)\sigma^a$ with '$x$' referring to the point which $l_i$ singles out in this limit. Here, $N(x)$ and $N^a(x)$ are seen to give the lapse and shift fields.  In (\ref{hmmmdi}) we have omitted certain zero-order terms and equation (\ref{hmmmdi}) is therefore seen to equal the principal part of the Dirac Hamiltonian in $3+1$ dimensions. Thus, the states (\ref{dadada}) can be interpreted as one-particle states on which the Dirac type operator $\cd_M$ gives the Hamiltonian in the semi-classical approximation.
We refer to \cite{AGNP1} for a more detailed discussion of this interpretation.

\section{ Many-Particle states}
\label{many}

In the states (\ref{sss}) the sum runs over parallel transports which start at the basepoint, travels up to a spinor degree of freedom (the matrix factor $\psi(x_i)$) at a vertex, and then back to the basepoint. The parallel transport going up to the spinor utilizes the operators ${\bf U}_p$ whereas the parallel transport going back uses the operators $U_p$, see (\ref{twooperators}). The two path {\it up} and {\it down} are in (\ref{sss}) taken to be identical, but one could also permit different paths. Thus, the spinor degrees of freedom arise together with a loop composed of ${\bf U}_p$ and $U_p$ operators and is located at the point where these are joined. 
In this section we analyze objects in $\ch$ which involves {\it products} of loops composed of ${\bf U}_p$'s and $U_p$'s. Each of these loops carry spinor degrees of freedom, and thus the number of spinor degrees of freedom will grow together with the number of loop factors. These loop objects will be seen to form a system of many-particle states in the semi-classical continuum limit.

We start the analysis by constructing states which are build from $\Psi_{n}(\psi) $ and the coherent states $\phi^t_n$.
Thus, we write down the anti-symmetrized state 
\begin{equation}
\Psi^t_{m,n}(\psi_1,\ldots,\psi_m,\phi^t_n):= \sum_{\sigma\in S_m} (-1)^{\vert\sigma\vert} \Psi_{n}(\psi_{\sigma(1)})\ldots\Psi_{n}(\psi_{\sigma(m)})\Big\vert_{\mbox{\small grad$\le1$}}\phi_n^t
\label{supermanN}
\end{equation}
where $S_m$ is the group of $m$ permutations and where "grad$\le1$" means that we only allow terms in this sum which involve at most one path ${\bf U}_{p_i}$ which is odd with respect to the Clifford algebra. Furthermore, we shall restrict the choice of paths $p_i$ in (\ref{supermanN}) so that the operator ${\bf U}_{p_i \cap p_j}$, which corresponds to the intersection $p_i \cap p_j$ between two paths, is even with respect to the Clifford algebra for all paths in the sum in (\ref{supermanN}). These restriction appears to be rather unnatural but will play a central role in the following analysis. We shall comment on this below.

Before we continue let us comment briefly on the structure of (\ref{supermanN}). When we construct states in $\ch$ which we wish to assign the prefix "physical" then gauge invariance must be taken into consideration. A priori, it is not obvious how one should define an action of the gauge group to the matrix factor in $\ch$ -- and thereby to the spinor degrees of freedom which turn up in (\ref{supermanN}) -- but in the continuum and semi-classical limit one must demand that only terms which are gauge covariant and invariant emerge (in this limit one can define the ordinary gauge transformation for the spinors, but it will not define an action of the gauge group on $\ch$). Therefore, physical states can only involve loops. This is why loops, build of $U_p$'s and ${\bf U}_p$'s, are natural object to consider.

We now proceed to compute both the inner product and the expectation value of $\cd$ on such states in the semi-classical approximation. In the following we shall not be concerned with the lapse and shift fields and therefore only work with the Dirac type operator $\cd$ (and not $\cd_M$).

\subsection{The two-particle sector}

Consider the case $m=2$ 
$$
\Psi^t_{2,n}(\psi_1,\psi_2,\phi^t_n)=2^{-6n} \sum_{i,j}\left({\bf U}_{p_i}\psi_1(x_i) U^{-1}_{p_i}{\bf U}_{p_j}\psi_2(x_j) U^{-1}_{p_j}\right)\phi^t_n - (\psi_1\leftrightarrow \psi_2)\;.
$$
A long and tedious computation, which is carried out in appendix A, gives first
\begin{eqnarray}
\langle \Psi_{2,n}(\psi_1,\psi_2,\phi^t_n)  \vert \Psi_{2,n}(\psi_1,\psi_2,\phi^t_n) \rangle
\hspace{-2cm}&&\nonumber\\
&=&
 \langle \Psi(\psi_2)\vert \Psi(\psi_2)\rangle\langle \Psi(\psi_1)\vert \Psi(\psi_1)\rangle
+ \langle \Psi(\psi_2)\vert \Psi(\psi_2)\rangle \langle \Psi(\psi_1)\vert \Psi(\psi_1)\rangle 
 \nonumber\\& &
+ \frac{2^{-6n}}{4} \sum_{ij} 
 \mbox{Tr}_{M_2}\left( U_{p_j} \psi^*_2(x_{j}) \psi_1(x_j) U^{-1}_{p_j} U_{p_i}\psi^*_1(x_i)\psi_2(x_i)U^{-1}_{p_i}\right)
\nonumber\\&& + \mbox{terms which vanish in the semi-classical continuum limit}
\nonumber\\&& + \mbox{terms from anti-symmetrization.}
\label{herefirst}
\end{eqnarray}
The first two terms in (\ref{herefirst}) are identical to expressions one would obtain from the one-particle sectors of $\psi_1$ and $\psi_2$ respectively. The last term, however, is a "cross-term" which involves parallel transports $U_p$'s. If we take the combined semi-classical and continuum limit of the last term we find
\begin{eqnarray}
\stackrel{\mbox{\tiny cl + cont.}}{\longrightarrow}  \frac{1}{4}\int_\Sigma dx \int_\Sigma dy&&\hspace{-2mm}
 \mbox{Tr}_{M_2}\big( U(y,x) \psi^*_2(x)  \psi_1(x) U(x,y) \psi^*_1(y)\psi_2(y)\big) 
\nonumber\\&&
 + \mbox{terms from anti-symmetrization,}
  \label{interactiontermdada}
\end{eqnarray}
where $U(x,y)=U^{-1}(y,x)$ is the parallel transport between points $x$ and $y$ in the continuum lattice, given by the composition of parallel transports $U_{p_i}$'s. Thus, in the continuum limit we find terms where two fermions are connected by a gravitational flux tube.
In the definition (\ref{sss}) of $\Psi_{n}(\psi)$ we could have included a normalized sum over paths connecting the basepoint with each matrix $\psi(x_i)$. In that case $U(x,y)$ in (\ref{interactionterm}) would involve a sum over paths connecting $x$ and $y$ and thus the interaction terms would involve several flux tubes. Notice also that although all $U_{p}$'s start at the basepoint the composition $U(x,y)$ does not necessarily run through the basepoint: two parallel transports $U_{p_i}$ and $U_{p_j}$ may have an overlap, $U_{p_j}\cap U_{p_i}\not=0$, so that their composition $U_{p_i}^*\cdot U_{p_j}$ will only be connected to the basepoint through a trivial backtracking which cancels out.

The computation of (\ref{herefirst}) depends crucially on the presence of the Clifford algebra elements in ${\bf U}_p$ and the fact that conjugation with ${\bf U}_p$ corresponds to a trace over the matrices, see (\ref{Trace}). 
The factor $\frac{1}{4}$ in (\ref{herefirst}) comes from a trace over $M_2$ that emerge from a conjugation with ${\bf U}_p$, as shown in appendix A.

Finally, in the particular limit where we let the coherent states $\phi_n^t$ be peaked over the classical phase space point which corresponds to a flat space-time with $U(x,y)\equiv1$, then this expression gives
\begin{eqnarray}
\stackrel{\mbox{\tiny flat-space}}{\longrightarrow}
 &=&
\frac{1}{4} \int_\Sigma dx\int_\Sigma dy \mbox{Tr}_{M_2}\big(  \psi^*_2(x)  \psi_1(x)   \psi^*_1(y)\psi_2(y)\big) 
\nonumber\\ &&
 + \mbox{terms from anti-symmetrization.}
 \label{flatspacelimitdA}
\end{eqnarray}
Ignoring the factor $\frac{1}{4}$ this term resemble a cross term coming from an anti-symmetrized Fock space. However, such a term would be of the form
\begin{equation}
 \int_\Sigma dx\mbox{Tr}\big(  \psi^*_2(x) \psi_1(x)\big)  \int_\Sigma dy \mbox{Tr}\big( \psi^*_1(y)\psi_2(y)\big) 
\label{sara}
\end{equation}
and we see that the difference is the occurrence of an additional trace. This means that the cross-term term (\ref{interactiontermdada}) can, in the flat-space limit, almost be interpreted in terms of anti-symmetrization of a fermionic Fock space, except that an odd mixing of spinor degrees of freedom happen due to the missing trace in (\ref{flatspacelimit}). We shall further analyze this issue in section \ref{analize}.

Let us now work out the expectation value of the Dirac type operator $\cd$ on the states (\ref{supermanN}). Again, the computations are given in appendix A. First, we find
\begin{eqnarray}
\langle \Psi_{2,n}(\psi_1,\psi_2,\phi^t_n) \vert \cd \vert \Psi_{2,n}(\psi_1,\psi_2,\phi^t_n) \rangle
\hspace{-6cm}&&\nonumber\\
&=&
 \langle \Psi(\psi_2)\vert \Psi(\psi_2)\rangle\langle \Psi(\psi_1)\vert \cd\vert \Psi(\psi_1)\rangle
+ \langle \Psi(\psi_2)\vert \cd\vert \Psi(\psi_2)\rangle \langle \Psi(\psi_1)\vert \Psi(\psi_1)\rangle 
 \nonumber\\& &
+ \frac{2^{-6n}}{4} \sum_{ij} 
\Big(
 \mbox{Tr}_{M_2}\left( U_{p_j} g_{j+1} \psi^*_2(x_{j+1}) g^*_{j+1} \mathrm{i} \sigma^a E^{m}_a  \psi_1(x_j) U^{-1}_{p_j} U_{p_i}\psi^*_1(x_i)\psi_2(x_i)U^{-1}_{p_i}\right)
\nonumber\\&&
-   \mbox{Tr}_{M_2}\left(  U_{p_j} \psi^*_2(x_j)\mathrm{i} \sigma^a E^{m}_a g_{j+1} \psi_1(x_{j+1})g_{j+1}^* U^{-1}_{p_j} U_{p_i}\psi^*_1(x_i)\psi_2(x_i)U^{-1}_{p_i}\right) 
\nonumber\\&&
+ \mbox{Tr}_{M_2}\left(  U_{p_j} \psi^*_2(x_j)\psi_1(x_j) U^{-1}_{p_j} U_{p_i}  g_{i+1}\psi^*_1(x_{i+1}) g^*_{i+1}\mathrm{i} \sigma^a E^{m}_a  \psi_2(x_i)U^{-1}_{p_i}\right)
\nonumber\\& &
-  \mbox{Tr}_{M_2}\left(  U_{p_j} \psi^*_2(x_j)\psi_1(x_j) U^{-1}_{p_j} U_{p_i}\psi^*_1(x_i) \mathrm{i} \sigma^a E^{m}_a g_{i+1} \psi_2(x_{i+1})g^*_{i+1} U^{-1}_{p_i}\right) \Big) 
\nonumber\\&& + \mbox{terms which vanish in the semi-classical continuum limit}
\nonumber\\&& + \mbox{terms from anti-symmetrization,}
\label{here}
\end{eqnarray}
where expressions like $E^{m}_a\sigma^a g_{j+1}\psi^*_2(x_{j+1}) g^*_{j+1}$ (sum over $a$) refer again to an edge which starts where $U_{p_j}$ ends and which has the matrix $\psi^*_2$ associated to its endpoint, with the direction of the edge denoted by '$m$'. 

The computation of (\ref{herefirst}) and (\ref{here}), which is found in appendix A (see (\ref{UDREGNING})), is highly sign sensitive. The special conventions which we introduced with (\ref{supermanN}), where we permit only one odd ${\bf U}_p$ in the product, and where we permit only even intersections of the paths $p_i$, are crucial for the result (\ref{here}). Had we not introduced these conventions we would not, in all terms, have obtained the right sign which gives the derivative. Instead terms would arise from (\ref{UDREGNING}) which would diverge in the semi-classical continuum limit because they would not involve a $2^{-n}$ factor.

The first two terms in (\ref{here}) are terms which give the one-particle Hamiltonians in the semi-classical limit for the spinors $\psi_1$ and $\psi_2$. 
The computation of the one-particle sector was worked out in the previous section. The last four terms look like an interaction between particles $\psi_1$ and $\psi_2$. If we take the continuum and semi-classical limit of these last terms, we find 
\begin{eqnarray}
\stackrel{\mbox{\tiny cl + cont.}}{\longrightarrow}   \frac{1}{4}
\int_\Sigma dx \int_\Sigma dy&&\hspace{-5mm}
\Big( \mbox{Tr}\big( U(y,x) \psi^*_2(x) \not\hspace{-1mm}\nabla \psi_1(x) U(x,y) \psi^*_1(y)\psi_2(y)\big) 
\nonumber\\ &-&
 \mbox{Tr}\big( U(y,x) \not\hspace{-1mm}\nabla \psi^*_2(x) \psi_1(x) U(x,y) \psi^*_1(y)\psi_2(y)\big) 
 \nonumber\\ &+&
 \mbox{Tr}\big( U(y,x) \psi^*_2(x) \psi_1(x) U(x,y)  \not\hspace{-1mm}\nabla\psi^*_1(y)\psi_2(y)\big) 
 \nonumber\\ &-&
  \mbox{Tr}\big( U(y,x)  \psi^*_2(x) \psi_1(x) U(x,y) \psi^*_1(y) \not\hspace{-1mm}\nabla\psi_2(y)\big) \Big)
  \nonumber\\&& + \mbox{terms from anti-symmetrization.}
  \label{interactionterm}
\end{eqnarray}
where $U(x,y)=U^{-1}(y,x)$ is the parallel transport between points $x$ and $y$ in the continuum lattice, given by the composition of parallel transports $U_{p_i}$'s. Thus, in the continuum limit we find interaction terms where two fermions interact locally with a non-local gravitational flux tube.

Finally, we again consider the particular limit where the coherent states $\phi_n^t$ are peaked over the classical phase space point which corresponds to a flat space-time with $U(x,y)\equiv1$ and $ \not\hspace{-1mm}\nabla \equiv \not\hspace{-1mm}\pa $, then this expression gives
\begin{eqnarray}
\stackrel{\mbox{\tiny flat-space}}{\longrightarrow}\sim
 &&
\frac{1}{2} \int_\Sigma dx\int_\Sigma dy \mbox{Tr}\big(  \psi^*_2(x) \not\hspace{-1mm}\pa \psi_1(x)   \psi^*_1(y)\psi_2(y)\big) 
\nonumber\\ &+&
\frac{1}{2} \int_\Sigma dx  \int_\Sigma dy \mbox{Tr}\big(   \psi^*_2(x) \psi_1(x) \psi^*_1(y) \not\hspace{-1mm}\pa \psi_2(y)\big)
 \nonumber\\&& + \mbox{terms from anti-symmetrization.}
 \label{flatspacelimit}
\end{eqnarray}
The previous discussion regarding similarities with a free fermionic quantum field theory also applies here. We will further analyze these terms
in the next section.

\subsection{The flat space-time limit: Weyl spinors}
\label{analize}

Let us examine the terms in (\ref{sara}) and (\ref{flatspacelimit}) which emerge in the limit where we approximate the construction around a flat space-time geometry. For simplicity, we consider terms without the derivative, thus coming from  (\ref{sara}). These terms have the form
\begin{eqnarray}
 \int_\Sigma dx\int_\Sigma dy \mbox{Tr}\big(  \psi^*(x)  \phi(x)   \phi^*(y)\psi(y)\big) \;,
 \nonumber
 \label{flatspacelimit2}
\end{eqnarray}
where $\psi(x)$ and $\phi(x)$ are fields which takes value in two-by-two matrices and which should be interpreted as spinor fields.
If we rewrite this expression in terms of Weyl spinors 
\begin{equation}
\psi = \big(\psi_1,\psi_2\big)\;,\quad \phi = \big(\phi_1,\phi_2\big)\;,
\label{expand}
\end{equation}
where $\psi_i,\phi_i$ are the columns in $\psi$ and $\phi$ which represent Weyl components, then (\ref{flatspacelimit2}) becomes
\begin{eqnarray}
  && \int_\Sigma   (\psi^*_1 \cdot \phi_1) \int_\Sigma   ( \phi^*_1 \cdot \psi_1)   + 
    \int_\Sigma   (\psi^*_2 \cdot \phi_2) \int_\Sigma   ( \phi^*_2 \cdot \psi_2)   +
    \nonumber\\&&
      \int_\Sigma   (\psi^*_1 \cdot \phi_2) \int_\Sigma   ( \phi^*_2 \cdot \psi_1)   + 
       \int_\Sigma   (\psi^*_2 \cdot \phi_1) \int_\Sigma   ( \phi^*_1 \cdot \psi_2)    \;.
    \label{oddterms}   
\end{eqnarray}
On the other hand, consider the case where $\phi$ and $\psi$ appear in a two-particle state in a fermionic Fock space and consider in particular the terms coming from the anti-symmetrization of this state:
\begin{eqnarray}
  && \int_\Sigma   (\psi^*_1 \cdot \phi_1) \int_\Sigma   ( \phi^*_1 \cdot \psi_1)   + 
    \int_\Sigma   (\psi^*_2 \cdot \phi_2) \int_\Sigma   ( \phi^*_2 \cdot \psi_2)   +
    \nonumber\\&&
     2 \int_\Sigma   (\psi^*_2 \cdot \phi_2) \int_\Sigma   ( \phi^*_1 \cdot \psi_1)   \;.
\label{antisymmm}
\end{eqnarray}
The two expressions (\ref{oddterms}) and (\ref{antisymmm}) are very similar, their difference being an odd mixing of Weyl components in the last two terms of (\ref{oddterms}) an the absence in  (\ref{oddterms})  of the last term in (\ref{antisymmm}). The mixing of Weyl components in (\ref{oddterms}) appears to be unphysical, but one notices that if we restrict our construction to describe Weyl spinors only (thus, letting one of each columns in (\ref{expand}) be zero), then  (\ref{oddterms}) and (\ref{antisymmm}) coincide. The result is then, due to anti-symmetrization, that the entire two-particle sector obtains an overall factor which will be killed by the normalization. This feature generalizes to many-particle states. Therefore we conclude that
\begin{enumerate}
\item
the construction coincides with a free fermionic quantum field theory in the flat-space limit if one restricts it to involve only Weyl spinors, and
\item
when we consider four-spinors an odd mixing of Weyl components appear.
\end{enumerate}

The restriction to Weyl spinors does not appear very natural. Rather, we suspect that additional structure should be introduced to the construction which causes (\ref{oddterms}) and (\ref{antisymmm}) to be equal. Further, one might speculate whether this is related to the fact that we are working with $SU(2)$ connections and not with complexified $SU(2)$ connections, the latter being the original Ashtekar connection. In \cite{AGNP1} we have already proposed the idea that complexified $SU(2)$ connections may be introduced in our framework via a doubling of the Hilbert space together with the introduction of a real structure related to a reality condition. It may be that such an extension -- implemented in a natural way -- will cause equation (\ref{oddterms}) and (\ref{antisymmm}) to be equal. If this is possible one might even wonder if the anti-symmetrization in (\ref{supermanN}) is necessary: possible the cross terms of the form (\ref{interactionterm}) will descent, in the flat space-time limit, to terms which amounts to a anti-symmetrization.

\subsection{Many particle states}

The computations of sectors which involve more than two particles are very similar to the computations from the two-particle sector. Some details are given in appendix A.6. In general, what we find is that the continuum semi-classical limit of the expectation value of an $n$-particle state will, in addition to the $n$-particle sector also give all the $n-k$-sectors ($k\in\{1,2,... n-1\}$) and will involve pairs of fermions connected by one flux tubes, see equation (\ref{fluuuux}). Again, in the flat space-time limit we find that by restricting the construction to Weyl spinors we obtain what amounts to a $n$-particle sector coming from an anti-symmetric Fock-space.

\subsection{A second type of semi-classical states}

From the computation of the expectation value of $D$ on the states (\ref{supermanN}) one is lead to consider whether other types of many-particle states can be found in $\ch$. 
In fact, with the infinite dimensional Clifford bundle in $\ch$ there is some room to play in. In the following we shall discuss one possibility.

Consider the two loops in $\G_n$ where one is build out of ${\bf U}$'s except for one edge, and the other is build entirely from ${\bf U}$'s. Both loops have a matrix degree of freedom $\psi$ inserted:
\begin{equation}
\X_1(\psi) = {\bf U}_{p_1} \psi(x_i) U_i {\bf U}_{p_2}\;,\quad \X_2(\psi) = {\bf U}_{p_1} {\bf U}_i  \psi(x_{i+1}){\bf U}_{p_2}
\label{elementofcrime}
\end{equation}
so that the composition $(p_1, l_i,p_2) $ form a based loop in $\cb_{\G_n}$ and where $x_i$ and $x_{i+1}$ are the vertices where $l_i$ starts and ends respectively. Thus, $\X_1$ involves a loop build out of ${\bf U}$'s and with a single "hole", and $\X_2$ involves the same loop, but without the hole.
If we take the expectation of $D$ on the sum $\X_1(\psi) + \X_2(\psi) $ we get again the basic building block of a spatial Dirac operator, corresponding to one of the term in (\ref{hmmmmm}). Thus, with a sum over different loops we can again recover expressions like (\ref{hmmm}) and (\ref{hmmmdi}).

The difference between this line of construction and what we did in the previous sections appears when we proceed to construct also many-particle states.  Either we do so by permitting products of loops, or we consider loops with more than one hole. In any case, the presence of ${\bf U}_p$'s "on both sides of" $\psi(x_i)$ in (\ref{elementofcrime}) has the effect that the parallel transports in (\ref{interactiontermdada}) and (\ref{interactionterm}) is absent and that instead a double trace appears. Thus, the whole issue with mixing of different Weyl spinor components disappear with this alternative construction. Therefore, with this approach it {is} possible to work with 4-spinors.

Thus, we find that there are at least two types of semi-classical states: the ones discussed in the previous sections, see (\ref{supermanN}), which entail a fermion interaction with gravitational flux tubes, and the ones described in this section, see (\ref{elementofcrime}), which come without this fermion interaction. Also, if we permit in (\ref{supermanN}) a normalized sum over several paths connecting the matrix $\psi(x_i)$ with the basepoint, then the result will be a fermion interaction which involves more than one gravitational flux tube. Whenever flux tubes are present an odd mixing of Weyl components appear in the flat space-time limit, a feature which is not present if we restrict the construction to Weyl spinors. Note also that
the states (\ref{supermanN}) and products of states (\ref{elementofcrime}) will span the same Hilbert space. The difference between the two types of states is solely the way the matrices $\psi(x_i)$ are inserted.

Further analysis is needed to understand better the form of these semi-classical states. It seems clear, however, that the Hilbert space $\ch$ should only be regarded as an intermediate structure which we for now use to construct the spectral triple. A more appropriate Hilbert space should most likely be found in the continuum limit as the Hilbert space spanned by states of the form (\ref{supermanN}) (probably with both symmetric and anti-symmetric sectors). One might speculate whether we are in fact recovering a GNS construction around the semi-classical states $\lim_{n\rightarrow\infty}\phi^t_n$.

\section{Discussion}

The key new insight presented in this paper is that central elements of fermionic quantum field theory can be {\it derived} from canonical quantum gravity in a semi-classical approximation by utilizing elements of noncommutative geometry.

Specifically, we show that an infinite system of interacting fermions emerges in a semi-classical approximation from a previously constructed semi-finite spectral triple over holonomy loops. Further, when all gravitational degrees of freedom are turned off a free fermionic quantum field theory emerge. 
This result is a strong indication that this model should be interpreted as describing quantized gravitational fields coupled to quantized matter fields. It is important to realize that the spectral triple construction does, a priori, only involve gravitational degrees of freedom. Therefore, matter degrees of freedom are, to some extend, emergent and canonical.

A further indication that this model describes quantized gravitational fields is that it encodes the kinematics of quantum gravity \cite{AGNP1}. This should be understood in the sense that the interaction between the algebra of holonomy loops and the Dirac type operator reproduces the structure of the Poisson bracket of general relativity formulated in terms of Ashtekar variables \cite{Ashtekar:1986yd,Ashtekar:1987gu}. Thus, the spectral triple is essentially a rearrangement of canonical quantum gravity cast in terms of Ashtekar variables. 

This paper is a continuation 
of a series of papers \cite{AGNP1,Aastrup:2010kb,Aastrup:2010ds,Aastrup:2009dy} devoted to the study of the 
semi-classical analysis of the spectral triple construction. Here, semi-classical refer to the gravitational fields and is computed in orders of the Planck length. In \cite{AGNP1} semi-classical states were constructed which, in a semi-classical approximation, entails the Dirac Hamiltonian in 3+1 dimensions. Thus, these states were interpreted as one-fermion states in a given foliation and a given background gravitational field. In this paper we recognize that these one-fermion states are the simplest examples of a much larger class of states which are labelled by a certain loop order. The one-particle states found in \cite{Aastrup:2010ds}  lie in the first order sector. At $n$'th order the states entail, in a semi-classical approximation, a system of $n$ interacting fermions. Depending on the exact way these states are constructed they may come with an interaction which involves flux-tubes of the Ashtekar connection. These flux tubes connect fermion degrees of freedom located at different points in space.
 Furthermore, we find that in the additional limit where all gravitational degrees of freedom are turned off a free fermionic quantum field theory emerge. Again, depending on the exact way the semi-classical states are constructed this free theory may involve a certain mixing of different spinor degrees of freedom. This mixing can be circumvented by restricting to Weyl spiors.\\

These new results raise many questions. First, one should clarify what happens to the free quantum field theory when one perturbs the background around the flat-space limit. The interesting question is what kind of interaction will emerge. One might speculate whether it will fall within the setup of an interacting quantum field theory. On a similar note, one should also analyze how the quantum corrections - which goes beyond the semi-classical approximation - will affect the free theory. Again, what interactions will arise? Will they be local and will they fall within the setup of an interacting quantum field theory?

Also, one need to determine whether the fermionic interaction which emerges from the semi-classical states is plausible as a physical interaction, just as one should aim to understand better the particular mixing of different Weyl components which comes from this interaction. The restriction to Weyl spinors which we identify as a possible way to circumvent this mixing does not appear natural. Rather, it seems that additional structure should be introduced to remedy this problem.

The computations leading to the many-particle sectors are highly sign-sensitive and depend on a complicated system of gradings related to lengths of paths. In fact, the final result relies on an apparently ad-hoc convention which eliminates alternating sign factors which would otherwise cause the result to diverge. Therefore, the sign-sensitivity does not appear natural and one might speculate whether some additional grading -- for instance in the form of a real structure -- could change this. 
Furthermore, the issue concerning the normalization of spinors discussed in \cite{AGNP1} is still unresolved and requires attention.

The semi-classical analysis presented in this paper operates with a double limit: first a semi-classical limit is taken at a finite level of discretization whereupon a continuum limit is taken. At the present level of analysis it is not known whether the continuum limit can be taken without the semi-classical approximation and what constraints such a limit might entail. This issue is likely to be related to that of diffeomorphisms. Where the spectral triple construction does not possess an apparent action of the diffeomorphism group the Hamiltonians which we obtain in the semi-classical continuum limit are invariant. Thus, diffeomorphisms turn up in the combined semi-classical continuum limit and it is therefore an interesting question what role they will play if one takes the continuum limit alone.  Further, it is important to note that the semi-classical approximation discussed in this paper is with respect to the Planck length $l_P$ and is a semi-classical limit of quantized gravitational fields. When elements of quantum field theory emerge in this limit the question arises how $\hbar$ might emerge from the construction. We suspect that this issue is related to the issue with the double limits.

The semi-classical states have a very particular form. They involve a sum over holonomy loops composed of two different kinds of  parallel transports: one is the ordinary one, the other involves also elements of the infinite dimensional Clifford algebra associated to the spectral triple construction. The latter resembles an $n$-form (at a finite level) and one might wonder what geometrical significance these states have, in relation to the spectral triple construction. Furthermore, spinor degrees of freedom appear through an insertion of matrices in these loops. These insertions appears somewhat arbitrary and one might wonder whether a guiding principle exist.  Also, in the construction of the semi-classical states we imposed an anti-symmetrization. Thus, a priori there will also be a symmetric sector in the Hilbert space. Whether this sector could somehow be related to bosonic degrees of freedom is an interesting question.

The spectral triple construction works for a compact Lie group, which we set equal to $SU(2)$. The Ashtekar connection, however, is in its original form a {\it complexified} $SU(2)$ connection, corresponding to $SL(2,\mathbb{C})$, which is non-compact. A choice of an $SU(2)$ connection corresponds either to a Euclidean setting or to a formulation where the Hamilton acquires an additional term. Clearly, it is desirable to extend the spectral triple construction to the complexified group and because we operate with a particular continuum limit which favors infinitesimal edges, we believe this might be done by doubling the Hilbert space and introducing a real structure.

Another interesting issue is the appearance of the lapse and shift fields in the construction. So far we have found several ways to introduce these fields via matrices associated to edges in the infinite lattice. It is, however, essential to understand what geometrical role -- in terms of the spectral triple construction -- these degrees of freedom play. If this is understood one might be lead to a natural formulation of a Wheeler-DeWitt equation.
 
Indeed, despite the advances made in this paper it should be stressed that the Hilbert space in the spectral triple construction can only be regarded as a kinematical Hilbert space. So far no Wheeler-DeWitt constraint has been constructed nor implemented. Thus, the analysis carried out in this paper should be interpreted as dealing with a fermionic sector in a larger yet to be understood framework which involves also a purely gravitational sector. In a forthcoming paper we will make a first step as to how such a gravitational sector can be found within this spectral triple framework.

\vspace{1cm}
\noindent{\bf Acknowledgements}\\
We are thankful to Ryszard Nest and Mario Paschke for numerous enlightning discussions. Also, we would like to thank Fedele Lizzi and Patrizia Vitale for hospitality and fruitful discussions during a visit.

\begin{appendix}
\section{Computations with many-particle states}

In this appendix we give first the computations concerning the properties of the operators ${\bf U}_p$ introduced in section \ref{dependency}. The bulk of this appendix gives the computations leading to the results (\ref{herefirst}) and (\ref{here}) in section \ref{many}.

\subsection{The operators ${\bf U}_p$}

We adopt the convention
$$(\fe_i^a)^*=-\fe_i^a \; \hbox{ and } \;(\sigma^a)^*=-\sigma^a\;.  $$
This convention is realized by
$$\sigma^1=\left( \begin{array}{cc}
0&-i\\
-i&0
\end{array} \right)\;, \quad \sigma^2=\left( \begin{array}{cc}
0&-1\\
1&0
\end{array} \right) \;,\quad \sigma^3= \left( \begin{array}{cc}
-\mathrm{i}&0\\
0&\mathrm{i}
\end{array} \right) \;. $$
With this realization we have 
$$\sigma^a \sigma^b =-\delta^{ab}+\epsilon^{abc}\sigma^c\;. $$
We begin by defining:
$$\cu_i=\frac{\mathrm{i}}2 \left( \fe_i^a \sigma^a+\fe_i^1 \fe_i^2 \fe_i^3  \right)\;.$$
The relationship between $\cu_i$ and ${\bf U}_i$ (which we introduce in section \ref{dependency}) is simply $\fU_i=\cu_i g_i$. We introduce here $\cu_i$ in order to keep the subsequent computations simpler.
Notice that $\cu_i$ is unitary since
\begin{eqnarray*} 
\cu_i \cu_i^*
&=&\frac{\mathrm{i}}2 \left( \fe_i^a \sigma^a+\fe_i^1 \fe_i^2 \fe_i^3  \right)  \frac{-\mathrm{i}}2 \left( \sigma^a \fe_i^a -\fe_i^3 \fe_i^2 \fe_i^1  \right) \\
&=& \frac14 \left( 4+ \sum_{a\not= b} \fe_i^a \sigma^a  \sigma^b \fe_i^b    -  \fe_i^a \sigma^a \fe_i^3 \fe_i^2 \fe_i^1   +  \fe_i^1 \fe_i^2 \fe_i^3     \sigma^a \fe_i^a    \right)\\
&=& 1+\frac14 \left( \epsilon_{abc}\sigma^a   \fe_i^b \fe_i^c  -\frac12 \epsilon_{abc}g_\mathrm{i}\sigma^a   \fe_i^b \fe_i^c -\frac12 \epsilon_{abc}\sigma^a   \fe_i^b \fe_i^c	\right)=1\;.
\end{eqnarray*}
Next we compute for $A \in M_2$ that
\begin{equation}  \label{spor}  Tr_{Cl} (\cu_i^* A\cu_i)= \frac14 \left(  \sigma^a \fe_i^a A \fe_i^a \sigma^a +   A  \right)=Tr_{M_2}(A)1_2, 
\end{equation}
since writing 
$$ A=a_01_2+a_a\sigma^a$$
and seeing that
$$ \sigma^a \fe_i^a \sigma^b \fe_i^a \sigma^a +   \sigma^b =0.$$
We will now make the additional convention $\fe_i^0:=\fe_i^1\fe_i^2\fe_i^3$ and $\sigma^0:=1_2$. The form of $\cu_i$ then reads 
$$\frac{\mathrm{i}}{2} \fe_i^a\sigma^a ,  $$
where the sum now runs over $a\in\{0,1,2,3\}$, and we furthermore have
$$\{ \fe_i^a , \fe_j^b \} =-2\eta^{ab}\;,$$
where $\eta$ is diagonal $\{-1,1,1,1\}$.

\subsection{The basic trick}
We will demonstrate the  basic trick for the computations we are going to do, by first computing the example
$$Tr ( A_n^*\cu_n^* \cdots A_1^* \cu_1^* B^*\cu^* \cu_1 C_1 \cdots \cu_n C_n \cu D ) .$$ 
We  note, that $\cu_1 C_1 \cdots \cu_n C_n$ is sandwiched between $\cu^*$ and $\cu$. We  write 
$$ \cu_1 C_1 \cdots \cu_n C_n=\frac{\mathrm{i}^n}{2^n}\fe_1^{a_1}\cdots \fe_n^{a_n}\sigma^{a_1}C_1 \cdots \sigma^{a_n} C_n.$$
Using (\ref{spor}) we get
\begin{eqnarray*}
\lefteqn{Tr ( A_n^*\cu_n^* \cdots A_1^* \cu_1^* B^*\cu^* \cu_1 C_1 \cdots \cu_n C_n \cu D )}\\
&=&Tr \left( A_n^*\cu_n^* \cdots A_1^* \cu_1^* B^*\cu^* \left(    \frac{\mathrm{i}^n}{2^n}\fe_1^{a_1}\cdots \fe_n^{a_n}\sigma^{a_1}C_1 \cdots \sigma^{a_n} C_n \right)  \cu D \right)\\
&=&(-1)^nTr \left( A_n^*\cu_n^* \cdots A_1^* \cu_1^* Tr_{M_2}\left(    \frac{\mathrm{i}^n}{2^n}\fe_1^{a_1}\cdots \fe_n^{a_n}\sigma^{a_1}C_1 \cdots \sigma^{a_n} C_n \right)B^* \cu^*   \cu D \right) \\
&=& (-1)^nTr \left( A_n^*\cu_n^* \cdots A_1^* \cu_1^* Tr_{M_2}\left(    \frac{\mathrm{i}^n}{2^n}\fe_1^{a_1}\cdots \fe_n^{a_n}\sigma^{a_1}C_1 \cdots \sigma^{a_n} C_n \right)B^* D \right) \;.
\end{eqnarray*}  
We next write 
$$Tr_{M_2}\left(    \frac{\mathrm{i}^n}{2^n}\fe_1^{a_1}\cdots \fe_n^{a_n}\sigma^{a_1}C_1 \cdots \sigma^{a_n} C_n \right) = \frac{\mathrm{i}}{2}\fe_1^a  \left(  \frac{\mathrm{i}^{n-1}}{2^{n-1}}\fe_2^{a_2}\cdots \fe_n^{a_n}C_1 \sigma^{a_2}C_2 \cdots \sigma^{a_n} C_n \right)^a   , $$
where we are summing over $a$ and  
$\left(  X \right)^a  $
means the $\sigma^a$-component of 
$X$.
Since 
$$\frac{\mathrm{i}^{n-1}}{2^{n-1}}\fe_2^{a_2}\cdots \fe_n^{a_n}\sigma^{a_2}C_2 \cdots \sigma^{a_n} C_n= C_1 \cu_2C_2 \cdots \cu_nC_n,$$
we will also write
$$\left(  \frac{\mathrm{i}^{n-1}}{2^{n-1}}\fe_2^{a_2}\cdots \fe_n^{a_n}C_1\sigma^{a_2}C_2 \cdots \sigma^{a_n} C_n \right)^a = \left( C_1 \cu_2C_2 \cdots \cu_nC_n \right)^a.$$
Continuing the computation we get
\begin{eqnarray*}
\lefteqn{Tr \left( A_n^*\cu_n^* \cdots A_1^* \cu_1^* Tr_{M_2}\left(    \frac{\mathrm{i}^n}{2^n}\fe_1^{a_1}\cdots \fe_n^{a_n}\sigma^{a_1}C_1 \cdots \sigma^{a_n} C_n \right)B^* D \right)  } \\
&=& Tr \left( A_n^*\cu_n^* \cdots A_1^* \left( \cu_1^* \frac{\mathrm{i}}{2}\fe_1^a \sigma^a \right)  \left( C_1 \cu_2C_2 \cdots \cu_nC_n \right)^a  B^* D \right) \\
&=& Tr \left( A_n^*\cu_n^* \cdots A_1^* \left(\left( \frac{\mathrm{i}}{2}\fe_1^b \sigma^b\right)^* \frac{\mathrm{i}}{2}\fe_1^a  \right)  \left( C_1 \cu_2C_2 \cdots \cu_nC_n \right)^a  B^* D \right)\\
&=& Tr \left( A_n^*\cu_n^* \cdots A_1^* \frac14  \left( C_1 \cu_2C_2 \cdots \cu_nC_n \right)  B^* D \right)\\
&=& \frac14 Tr(A_n^*C_n B^*D)Tr(A_1^*C_1)\cdots Tr(A_{n-1}C_{n-1})\;.
\end{eqnarray*}
All together we get
\begin{eqnarray*}
\lefteqn{Tr ( A_n^*\cu_n^* \cdots A_1^* \cu_1^* B^*\cu^* \cu_1 C_1 \cdots \cu_n C_n \cu D ) }\\
&=& \frac{(-1)^n}{4} Tr(A_n^*C_n B^*D)Tr(A_1^*C_1)\cdots Tr(A_{n-1}C_{n-1})\;.
\end{eqnarray*}
What is fundamental in this computation is the following (which we will use subsequently): We want to compute a term of the form
$$Tr( \tilde{A} \cu_1^* \tilde{B} \cu^*  \cu_1 \tilde{C} \cu \tilde{D}),$$
where $\tilde{A}, \tilde{B}, \tilde{C} ,\tilde{D}$ are almost general elements (see below) in $Cl(T^*_{id}\overline{\mathcal{A}})\otimes M_2$.
We can compute this as 
$$\frac{(-1)^\#}{4}(\tilde{A} \tilde{C}\tilde{B}\cu^*\cu \tilde{D}) =\frac{(-1)^\#}{4}(\tilde{A} \tilde{C} \tilde{B} \tilde{D}),$$
where $(-1)^\#$ is the sign we pick up by commuting the Clifford part of $\cu_1 \tilde{A}$ past the Clifford part of $\tilde{B} \cu^*$, i.e. $\#$ is the Clifford degree of $\cu_1 \tilde{C}$ multiplied with the Clifford degree of $\tilde{B} \cu^*$.  
That $\tilde{A}, \tilde{B}, \tilde{C} ,\tilde{D}$  are almost general means, that they dont have any Clifford elements in common with $\cu_1$ and $\cu$, and that $ \tilde{B}, \tilde{C} $ don't have any Clifford elements in common.

We can also remark, that if we replace  $\cu_1 \tilde{C}$ with $\tilde{C}_1 \cu_1 \tilde{C}_2$, then, since we are taking the $Tr_{M_2}$ in the computation, we can compute  
$$Tr( \tilde{A} \cu_1^* \tilde{B} \cu^*\tilde{C}_1  \cu_1 \tilde{C}_2 \tilde{C} \cu \tilde{D})=\frac{(-1)^\#}{4}Tr(\tilde{A}  \tilde{C_2}\tilde{C_1} \tilde{B} \tilde{D}), $$
where we are picking up an extra sign, from commuting the Clifford part of  $\tilde{C}_1$ with the Clifford part of $\cu_1\tilde{C}_2$.

\subsection{Terms appearing in the inner product}
When computing the norm of a two particle state, there appear terms of the following form
\begin{eqnarray} \label{diagonal}
&Tr( A^* \cu_{p_3}^* \cu_{p_1}^* B^*\cu_{p_2}^*\cu^*_{p_1} \cu_{p_1}\cu_{p_2} C\cu_{p_1}\cu_{p_3} D ) \;,\\
 & \label{ikkediagonal} Tr( A^*\cu^*_{p_2} \cu_{p_1}^*B^*\cu_{p_3}^* \cu_{p_1}^*    \cu_{p_1} \cu_{p_2} C\cu_{p_1} \cu_{p_3} D   )\;,\\
 & \label{skidt}Tr ( A^* \cu_{p_2}^*\cu_{p_1}^* B^*\cu_{p_2}^* \cu_{p_1}^* \cu_{p_1}\cu_{p_3} C \cu_{p_1} \cu_{p_3} D )\;,
\end{eqnarray}
where $p_1,p_2,p_3$ denote paths.

The terms (\ref{diagonal}) are easily computed
\begin{equation} \label{diares}
Tr( A^* \cu_{p_3}^* \cu_{p_1}^* B^*\cu_{p_2}^*\cu^*_{p_1} \cu_{p_1}\cu_{p_2} C\cu_{p_1}\cu_{p_3} D )=Tr(A^*D)Tr(B^*C). 
\end{equation}

The  terms (\ref{ikkediagonal}) can be computed as follows
\begin{eqnarray}
\lefteqn{Tr( A^*\cu^*_{p_2} \cu_{p_1}^*B^*\cu_{p_3}^* \cu_{p_1}^*    \cu_{p_1} \cu_{p_2} C\cu_{p_1} \cu_{p_3} D   )} \nonumber \\
&=&Tr\left( A^*\cu^*_{p_2} \cu_{p_1}^*B^*\cu_{p_3}^* \cu_{p_1}^*    \cu_{p_1} \cu_{p_2} C\cu_{p_1} \cu_{p_3} D   \right)\nonumber \\
&=&Tr \left( A^*\cu^*_{p_2} \cu_{p_1}^*B^*\cu_{p_3}^*  \cu_{p_2} C\cu_{p_1} \cu_{p_3} D  \right)  \nonumber \\
&=&Tr \left(( A^*\cu^*_{p_2} \cu_{p_1}^*B^*)\cu_{p_3}^*  (\cu_{p_2} C\cu_{p_1}) \cu_{p_3} D  \right)  \nonumber \\
&=& \frac{(-1)^{n_1n_2+(n_1+n_2)n_3} }{4}Tr(A^* \cu_{p_2}^*\cu_{p_2}C B^*\cu_{p_3}^*\cu_{p_3}D   ) \nonumber \\
&=& \label{ikkediares}\frac{(-1)^{n_1n_2+n_1n_3+n_2n_3} }{4}Tr(A^* C B^*D   ),
\end{eqnarray} 
where $n_1,n_2,n_3$ are the length of $p_1,p_2,p_3$.

To compute the terms (\ref{skidt}), we first write
\begin{eqnarray*}
\lefteqn{Tr A^* \cu_{p_2}^*\cu_{p_1}^* B^*\cu_{p_2}^* \cu_{p_1}^* \cu_{p_1}\cu_{p_3} C \cu_{p_1} \cu_{p_3} D ) }\\
&=& Tr(A^* \cu_{p_2}^*\cu_{p_1}^* B^*\cu_{p_2}^* \cu_{p_3} C \cu_{p_1} \cu_{p_3} D)
\end{eqnarray*}
To proceed we note that $\cu_i^*=-\cu_i$. We write 
$$ \cu_{p_3}=\cu_{p_{31}}\cu_{n_3}$$
where $p_{31}$ is $p_3$ minus the last edge. With this notation we get
\begin{eqnarray*}
\lefteqn{Tr(A^* \cu_{p_2}^*\cu_{p_1}^* B^*\cu_{p_2}^* \cu_{p_3} C \cu_{p_1} \cu_{p_3} D)}\\
&=& -Tr((A^* \cu_{p_2}^*\cu_{p_1}^* B^*\cu_{p_2}^* \cu_{p_{31}})\cu_{n_3}^*( C \cu_{p_1} \cu_{p_{31}})\cu_{n_3} D)\\
&=& \frac{(-1)^{(n_1+n_3-1)(n_3-1+n_2)+1}}{4}Tr(A^* \cu_{p_2}^*(\cu_{p_1}^*   \cu_{p_1}) \cu_{p_{31}}CB^*\cu_{p_2}^* \cu_{p_{31}}\cu_{n_3}^*\cu_{n_3} D) \\
&=& \frac{(-1)^{(n_1+n_3-1)(n_3-1+n_2)+1}}{4}Tr(A^* \cu_{p_2}^*  \cu_{p_{31}}CB^*\cu_{p_2}^* \cu_{p_{31}} D).
\end{eqnarray*}
We are therefore left with computing terms of the form
$$ Tr(A\cu_{1}\cdots \cu_{n}B \cu_1\cdots \cu_nC).$$
These can be computed in the following way:
\begin{eqnarray*}
\lefteqn{Tr(A\cu_{1}\cdots \cu_{n}B \cu_1\cdots \cu_nC)}\\
&=& Tr(A\cu_{1}\cdots \cu_{n}B \cu_1^* \cu_2 \cdots \cu_{n-1} \cu_n^*C)\\
&=& \frac{(-1)^{n-1}}{4}Tr (A \cu_1  (\cu_1^*\cu_{2}\cdots \cu_{n-1}B) \cu_2 \cdots \cu_n \cu_n^*C )\\
&=& \frac{(-1)^{n-1}}{4}Tr (A \cu_{2}\cdots \cu_{n-1}B \cu_2\cdots \cu_{n-1}C )\\
&&\vdots \\
&=& \left\{
\begin{array}{ll} 
\frac{(-1)^{(n-1)+(n-3)+\ldots 1 }}{4^{\frac{n}{2}}}Tr(ABC) & ,\quad n\hbox{ even} \\
\frac{1}{4^{\frac{n-1}{2}}} Tr(AC)Tr(B)& ,\quad n\hbox{ odd}\;.
\end{array}
\right.
\end{eqnarray*}
Thus, plugging in, we finally get
\begin{eqnarray}
\lefteqn{Tr( A^* \cu_{p_2}^*\cu_{p_1}^* B^*\cu_{p_2}^* \cu_{p_1}^* \cu_{p_1}\cu_{p_3} C \cu_{p_1} \cu_{p_3} D ) }  \nonumber \\
&=& 
\left\{
\begin{array}{ll}
 \frac{(-1)^{(n_1+n_3-1)(n_3-1+n_2)+1}}{4} \frac{(-1)^{(n_2+n_3)+(n_2+n_3-2)+\ldots 1 }}{4^{\frac{n_2+n_3}{2}}}
Tr(A^*CB^*D)&,\quad  n_2+n_3 \hbox{ odd} \\
\frac{(-1)^{(n_1+n_3-1)(n_3-1+n_2)+1}}{4}  \frac{1}{4^{\frac{n_2+n_3}{2}}} Tr(A^*D)Tr(CB^*) & , \quad n_2+n_3 \hbox{ even}\;. \label{skidtpar}
\end{array}
\right.
\end{eqnarray}
Notice here the factor 
$$
4^{\frac{n_2+n_3}{2}}
$$
appearing in the denominator of (\ref{skidtpar}). Thus, when $n_i\rightarrow\infty$ (the continuum limit) these terms vanish. This is very fortunate since such terms would not fit with the results of our computations.

\subsection{Acting with the Dirac type operator}
The corresponding terms for the Dirac operator $\cd$, which we need to compute, are roughly of the form
\begin{eqnarray} \label{ddiagonal}
& Tr( A^* \cu_{p_3}^* \cu_{p_1}^* B^*\cu_{p_2}^*\cu^*_{p_1}\cd (\cu_{p_1}\cu_{p_2} C\cu_{p_1}\cu_{p_3} D ))\;, \\
 & \label{ikkeddiagonal} Tr( A^*\cu^*_{p_2} \cu_{p_1}^*B^*\cu_{p_3}^* \cu_{p_1}^*\cd (    \cu_{p_1} \cu_{p_2} C\cu_{p_1} \cu_{p_3} D  ))\;, \\
 & \label{skidtd}Tr( A^* \cu_{p_2}^*\cu_{p_1}^* B^*\cu_{p_2}^* \cu_{p_1}^*\cd (\cu_{p_1}\cu_{p_3} C \cu_{p_1} \cu_{p_3} D ))\;.
\end{eqnarray}
Of course either $p_2$ or $p_3$ need be one edge shorter on one of the sides in order give something different from zero, and we also need to multiply with Hall coherent states, but we will omit this in the notation here.
 
For the computation of (\ref{ddiagonal}) let us first compute the term
$$  Tr( A^* \cu_{p_3}^* \cu_{p_1}^* B^*\cu_{p_2}^*\cu^*_{p_1}\cd (\cu_{p_1}\cu_{p_2} \cu_i C\cu_{p_1}\cu_{p_3} D ))  .$$ 
  This only gives something, for the term of the form 
  in $\cd$ which include $\fe_i^a$. Multiplying $\fe_i^a$ on $\cu_i$ and remembering that we are taking the trace at the end, gives $\frac{-\mathrm{i}}{2} \sigma^a$. We therefore get
\begin{eqnarray}
\lefteqn{Tr( A^* \cu_{p_3}^* \cu_{p_1}^* B^*\cu_{p_2}^*\cu^*_{p_1}\cd (\cu_{p_1}\cu_{p_2} \cu_j C\cu_{p_1}\cu_{p_3} D )) } \nonumber   \\
&=& (-1)^{n_1+n_2}Tr( A^* \cu_{p_3}^* \cu_{p_1}^* B^*\cu_{p_2}^*\cu^*_{p_1}\cu_{p_1}\cu_{p_2} \mathrm{i} E_a^j\frac{-\mathrm{i}}{2} \sigma^a C\cu_{p_1}\cu_{p_3} D ) \nonumber  \\
&=& (-1)^{n_1+n_2}Tr( A^*D) Tr( B^*  \mathrm{i} E_a^j\frac{-\mathrm{i}}{2} \sigma^a C) \;.\label{d1}
\end{eqnarray}
Here we have been  sloppy with the notation, and omitting the prefactor $(-1)^n$. $\mathrm{i}E_a^j$ denotes the expectation value of $L_{e_j^a}$ on semi-classical states, see (\ref{JP1}).

 Similarly, if the $\cu_i$'s are on the other side we get 
 \begin{equation} (-1)^{n_1+n_2}Tr( A^*D) Tr( B^* \mathrm{i}E_a^j\frac{\mathrm{i}}{2} \sigma^a C)\;. \label{d2} \end{equation}
 If we had prolonged $\cu_{p_3}$ instead we get
 \begin{equation} (-1)^{n_2+n_3}Tr( A^* \mathrm{i}E_a^j\frac{-\mathrm{i}}{2} \sigma^a D) Tr( B^*  C)    \label{d3}\end{equation} and 
\begin{equation} (-1)^{n_2+n_3}Tr( A^* \mathrm{i}E_a^j \frac{\mathrm{i}}{2} \sigma^a  D) Tr( B^*C) \label{d4} .\end{equation} 
  
Similarly for (\ref{ikkeddiagonal}) we get the terms (right hand side indicating where the $\cu_j$ has been inserted)
\begin{eqnarray}
\begin{array}{lll}
 & (-1)^{n_1+n_2+n_1n_2+n_1n_3+n_2n_3} Tr(A^*\mathrm{i}E_a^j\frac{-\mathrm{i}}{2} \sigma^a CB^*D  )\qquad&\big[ \cu_{p_2}\cu_jC   \label{id1}\big]\\
 & (-1)^{n_2+n_3+n_1n_2+n_1n_3+n_2n_3} Tr(A^*\mathrm{i}E_a^j\frac{\mathrm{i}}{2}  \sigma^a CB^*D  ) &\big[ A^*\cu_j^*\cu_{p_2}^*  \label{id2}\big]\\
 & (-1)^{n_2+n_3+n_1n_2+n_1n_3+n_2n_3} Tr(A^* CB^* \mathrm{i}E_a^j\frac{-\mathrm{i}}{2} \sigma^a D  )&\big[ \cu_{p_3}\cu_j D  \label{id3} \big]\\
 & (-1)^{n_1+n_3+n_1n_2+n_1n_3+n_2n_3} Tr(A^* CB^* \mathrm{i}E_a^j\frac{\mathrm{i}}{2} \sigma^a D  )&\big[ B^*\cu_j^* \cu_{p_3}^*\big]\;. \label{id4}
\end{array}
\end{eqnarray}
Similarly we could write the terms of the type (\ref{skidtd}). However in the continuum limit these terms will vanish, due to the powers of $\frac14$ appearing in these terms.
  
\subsubsection{Replacing the $\cu$'s with the $\fU$'s}
In order to keep the notation in the above calculations simple, we have done the entire computations with $\cu_i$ and not $\fU_i=\cu_i g_i$. The computations with $\fU_i$ instead of $\cu_i$ can  be done similarly. The results  (\ref{diares}), (\ref{ikkediares}), (\ref{d1},\ref{d2},\ref{d3},\ref{d4}), and (\ref{id1},\ref{id2},\ref{id3},\ref{id1})  remain the same. The reason for this is  that the $g$'s always cancels out with the $g^{-1}$'s. The results (\ref{skidtpar}) basically remain the same, however  the matrices $A,B,C,D$ will be multiplied by suitable $g$'es, or traces thereof. The powers of $\frac14$ remain the same, and the terms will therefore vanish in the continuum limit.

\subsection{Computation applied the two-particle states}
We will now apply the computation to the two particle states
\begin{equation}
 \sum_{i,j}\fU_{p_j} \psi_1 (x_j)U_{p_j}^{-1}  \fU_{p_i}\psi_2 (x_i)U_{p_i}^{-1}.
\label{exxxxx}
\end{equation}
We have here omitted the prefactor and the Hall's coherent state in the  notation. Also, we will give the computations for combinations of ${\bf U}_p$'s with arbitrary gradings, and thus not invoke the convention introduced in (\ref{supermanN}) that at most one ${\bf U}_p$ can be odd with respect to the Clifford algebra. The reason for this is that we wish to display the sign-sensititity of these computations. 
Finally, note that we have not antisymmetrized the expression (\ref{exxxxx}). Instead we compute the inner product with a term of the form 
$$ \sum_{i,j}\fU_{p_j} \psi_3 (x_j)U_{p_j}^{-1}  \fU_{p_i}\psi_4 (x_i)U_{p_i}^{-1}.$$
We see that three types of terms can arise in the inner product:
\begin{eqnarray}
 & \label{diacont} Tr(   U_{p_i} \psi_2 (x_i)^* \fU_{p_i}^* U_{p_j} \psi_1(x_j)^*\fU_{p_j}^*(  \fU_{p_j} \psi_3 (x_j)U_{p_j}^{-1}  \fU_{p_i}\psi_4 (x_i)U_{p_i}^{-1}   )) \\
& \label{offdiacont}Tr(   U_{p_i} \psi_2 (x_i)^* \fU_{p_i}^* U_{p_j} \psi_1(x_j)^*\fU_{p_j}^*(  \fU_{p_i} \psi_3 (x_i)U_{p_i}^{-1}  \fU_{p_j}\psi_4 (x_j)U_{p_j}^{-1} )) \\
 &Tr(    U_{p_i} \psi_2 (x_i)^* \fU_{p_i}^* U_{p_i} \psi_1(x_i)^*\fU_{p_i}^*(  \fU_{p_j} \psi_3 (x_j)U_{p_j}^{-1}  \fU_{p_j}\psi_4 (x_j)U_{p_j}^{-1} ))\;. \label{skidtcont}
\end{eqnarray}
According to (\ref{diares}), the terms (\ref{diacont}) gives 
$$ Tr( \psi_1(x_j)^*\psi_3(x_j))Tr(\psi_2(x_i)^*\psi_4(x_i))\;.$$
According to (\ref{ikkediares}), the terms (\ref{offdiacont}) gives
$$ \frac{(-1)^{nn_j+nn_i+n_jn_i} }{4}Tr(U_{p_i} \psi_2(x_i)^*\psi_3(x_i) U_{p_i}^{-1} U_{p_j}\psi_1(x_j)^* \psi_4 (x_j)U_{p_j}^{-1}) ,$$
where $n$ denotes the lenght of the path $p_j$ and $p_i$ have in common, $n_j$ the length of $p_j$ minus $n$ and $n_i$ the length of $p_i$ minus $n$.

The terms (\ref{skidtcont}) vanishes in the continuum limit due to the powers of $\frac14$ appearing in (\ref{skidtpar}).

\subsubsection{Expectation value of the Dirac operator}

We will now compute the terms (\ref{ddiagonal}), (\ref{ikkeddiagonal}) and (\ref{skidtd}) for the two particle states. Again each of these terms basically constists of 4 terms, in the case (\ref{ddiagonal}) these terms are
\begin{eqnarray*}
& Tr(    U_{p_i} \psi_2 (x_i)^* \fU_{p_i}^* U_{p_j} \psi_1(x_j)^*\fU_{p_j}^* \cd ( \fU_{p_j} \fU_{j+1} \psi_3 (x_{j+1})U_{p_{j+1}}^{-1}  \fU_{p_i}\psi_4 (x_i)U_{p_i}^{-1}  )) \\
& Tr(    U_{p_i} \psi_2 (x_i)^* \fU_{p_i}^* U_{p_{j+1}} \psi_1(x_{j+1})^*  \fU_{j+1}^*\fU_{p_j}^* \cd (\fU_{p_j} \psi_3 (x_j)U_{p_j}^{-1}  \fU_{p_i}\psi_4 (x_i)U_{p_i}^{-1}  )) \\
& Tr(   U_{p_i} \psi_2 (x_i)^* \fU_{p_i}^* U_{p_j} \psi_1(x_j)^*\fU_{p_j}^*\cd ( \fU_{p_j} \psi_3 (x_j)U_{p_j}^{-1}  \fU_{p_i}\fU_{i+1}\psi_4 (x_{i+1})U_{p_{i+1}}^{-1}   )) \\
& Tr(   U_{p_{i+1}} \psi_2 (x_{i+1})^*\fU_{i+1} \fU_{p_i}^* U_{p_j} \psi_1(x_j)^*\fU_{p_j}^* \cd (\fU_{p_j} \psi_3 (x_j)U_{p_j}^{-1}  \fU_{p_i}\psi_4 (x_i)U_{p_i}^{-1}  )) 
\end{eqnarray*}
where the $i+1$ denotes going one edge further than $p_i$, and $x_{i+1}$ denotes the endpoint of this edge. The terms give
\begin{eqnarray*}
&-\frac12 Tr(\psi_1(x_j)^* iE_a^{j+1}(x_{j+1})\mathrm{i}\sigma^a g_{j+1} \psi_3(x_{j+1})g_{j+1}^{-1})Tr(\psi_2(x_i)^*\psi_4(x_i))\\
&\frac12 Tr(g_{j+1}\psi_1(x_{j+1})^* g_{j+1}^{-1} iE_a^{j+1}(x_{j+1})\mathrm{i}\sigma^a\psi_3(x_j))Tr(\psi_2(x_i)^*\psi_4(x_i))\\
&(-1)^{n_j+n_i+1}\frac12 Tr(\psi_1(x_j)^* \psi_3(x_{j}))Tr(\psi_2(x_i)^*  i E_a^{i+1}(x_{i+1})\mathrm{i}\sigma^a g_{i+1}\psi_4(x_{i+1})g_{i+1}^{-1})\\
&(-1)^{n_j+n_i}\frac12 Tr(\psi_1(x_j)^* \psi_3(x_{j}))Tr( g_{i+1}\psi_2(x_{i+1})^*g_{i+1}^{-1}  i E_a^{i+1}(x_{i+1})\mathrm{i}\sigma^a \psi_4(x_i))
\end{eqnarray*}
where we have the same conventions for $n,n_j,n_i$ like in the computation of the inner product of the two particle state, and where $g_{i+1}$ means the copy of $G$ associated to the $i+1$-edge.

Similarly the terms of (\ref{ikkeddiagonal}) give
\begin{eqnarray}
& \frac{(-1)^{1+n+n_i+nn_j+nn_i+n_1n_i}}{4}   Tr( U_{p_i}      \psi_2(x_i)^*    i E_a^{i+1}(x_{i+1})\mathrm{i}\sigma^a g_{i+1}\psi_3(x_{i+1})g_{i+1}^{-1}U_{p_i}^{-1}U_{p_j}   \psi_1 (x_j)^*\psi_4(x_j)U_{p_j}^{-1})\nn\\
&\frac{(-1)^{n_j+n_i+nn_j+nn_i+n_1n_i}}{4} Tr(   U_{p_i}g_{i+1}     \psi_2(x_{i+1})^* g_{i+1}^{-1}   i E_a^{i+1}(x_i)\mathrm{i}\sigma^a \psi_3(x_{i})U_{p_i}^{-1}U_{p_j}\psi_1 (x_j)^*\psi_4(x_j)U_{p_j}^ {-1})\nn\\
& \frac{(-1)^{1+n_j+n_i+nn_j+nn_i+n_jn_i}}{4}Tr(    U_{p_i}   \psi_2(x_{i})^*     \psi_3(x_{i})U_{p_i}^ {-1} U_{p_j}\psi_1 (x_j)^* iE_a^{j+1}(x_{j+1})\mathrm{i}\sigma^a   g_{j+1}^{-1} \psi_4(x_{j+1}) g_{j+1}^{-1}U_{p_j}^{-1})\nn\\
& \frac{(-1)^{n+n_j+nn_j+nn_i+n_jn_i}}{4}Tr(  U_{p_i}     \psi_2(x_{i})^*     \psi_3(x_{i})U_{p_i}^{-1}U_{p_j}g_{j+1}\psi_1 (x_{j+1})^* g_{j+1}^{-1} i E_a^{j+1}(x_{j+1})\mathrm{i}\sigma^a    \psi_4(x_{j}) U_{p_j}^{-1})
\nn\\
\label{UDREGNING}
\end{eqnarray}
where we again have the same conventions for $n,n_j,n_i$ like in the computation of the inner product of the two particle state.

Again terms of the form (\ref{skidtd}) will wanish in the continuum limit.

\subsection{$n\ge3$ particle states}

As we can see for the two-particle states, the computations are somewhat involved, so we will only indicate how the computations can be done for many particle states. 

The computations for many particles reduces to the problem of computing terms of the form
$$ Tr( \psi^*_{1n}\fU_{p_n}^*\cdots \psi^*_{11} \fU_{p_1}^*\fU_{p_{\sigma (1)}}\psi_{2\sigma (1)}\cdots \fU_{p_{\sigma (n)}} \psi_{2\sigma (n)}    )\; ,   $$
where $\sigma$ is a permutation of $\{ 1,\ldots ,n\}$. 
The computations can be done similar to those of the two particle states. One starts by finding the first $\fU_{p_1}$ in $$\fU_{p_{\sigma (1)}}\psi_{2\sigma (1)}\cdots \fU_{p_{\sigma (n)}} \psi_{2\sigma (n)}\;. $$ 
One then moves the expression, which is sandwiched between $\fU_{p_1}$ and $\fU_{p_1}^*$ to the first $\fU_{p_{\sigma(1)}}^*$ appearing in 
$$ \psi^*_{1n}\fU_{p_n}^*\cdots \psi^*_{11} \fU_{p_1}^*\;, $$
and cancels out $\fU_{p_{\sigma (1)}}$ with $\fU_{p_{\sigma (1)}}^*$, and $\fU_{p_1}$ with $\fU_{p_1}^*$. One is now left with a nested expression of the same sort as one started with. One can therefore continue the procedure until all the $\fU_{p_i}$'s have been cancelled. 
Eventually, one will end up with terms, which are products of  
\begin{equation}
Tr(\psi_{1i}^*\psi_{1 i})\hbox{ or } Tr(U_{p_i}\psi_{1i}^*\psi_{2i}U^{-1}_{p_{i}} U_{p_j} \psi_{1j}^*\psi_{2j} U_{p_j}^{-1})   
\label{fluuuux}
\end{equation}
and of course  a prefactor, which is a power of $\frac14$ and a sign.

\end{appendix}

\end{document}